\PassOptionsToPackage{unicode}{hyperref}
\PassOptionsToPackage{hyphens}{url}
\PassOptionsToPackage{dvipsnames,svgnames,x11names}{xcolor}
\documentclass[
  12pt]{article}

\usepackage{amsmath,amssymb}
\usepackage{iftex}
\ifPDFTeX
  \usepackage[T1]{fontenc}
  \usepackage[utf8]{inputenc}
  \usepackage{textcomp} 
\else 
  \usepackage{unicode-math}
  \defaultfontfeatures{Scale=MatchLowercase}
  \defaultfontfeatures[\rmfamily]{Ligatures=TeX,Scale=1}
\fi
\usepackage{lmodern}
\ifPDFTeX\else  
\fi
\IfFileExists{upquote.sty}{\usepackage{upquote}}{}
\IfFileExists{microtype.sty}{
  \usepackage[]{microtype}
  \UseMicrotypeSet[protrusion]{basicmath} 
}{}
\makeatletter
\@ifundefined{KOMAClassName}{
  \IfFileExists{parskip.sty}{%
    \usepackage{parskip}
  }{
    \setlength{\parindent}{0pt}
    \setlength{\parskip}{6pt plus 2pt minus 1pt}}
}{
  \KOMAoptions{parskip=half}}
\makeatother
\usepackage{xcolor}
\makeatletter
\ifx\paragraph\undefined\else
  \let\oldparagraph\paragraph
  \renewcommand{\paragraph}{
    \@ifstar
      \xxxParagraphStar
      \xxxParagraphNoStar
  }
  \newcommand{\xxxParagraphStar}[1]{\oldparagraph*{#1}\mbox{}}
  \newcommand{\xxxParagraphNoStar}[1]{\oldparagraph{#1}\mbox{}}
\fi
\ifx\subparagraph\undefined\else
  \let\oldsubparagraph\subparagraph
  \renewcommand{\subparagraph}{
    \@ifstar
      \xxxSubParagraphStar
      \xxxSubParagraphNoStar
  }
  \newcommand{\xxxSubParagraphStar}[1]{\oldsubparagraph*{#1}\mbox{}}
  \newcommand{\xxxSubParagraphNoStar}[1]{\oldsubparagraph{#1}\mbox{}}
\fi
\makeatother

\usepackage{longtable,booktabs,array}
\usepackage{calc} 
\usepackage{etoolbox}
\makeatletter
\patchcmd\longtable{\par}{\if@noskipsec\mbox{}\fi\par}{}{}
\makeatother
\IfFileExists{footnotehyper.sty}{\usepackage{footnotehyper}}{\usepackage{footnote}}
\makesavenoteenv{longtable}
\usepackage{graphicx}
\makeatletter
\def\maxwidth{\ifdim\Gin@nat@width>\linewidth\linewidth\else\Gin@nat@width\fi}
\def\maxheight{\ifdim\Gin@nat@height>\textheight\textheight\else\Gin@nat@height\fi}
\makeatother
\setkeys{Gin}{width=\maxwidth,height=\maxheight,keepaspectratio}
\makeatletter
\def\fps@figure{htbp}
\makeatother

\makeatletter
\@ifpackageloaded{caption}{}{\usepackage{caption}}
\AtBeginDocument{%
\ifdefined\contentsname
  \renewcommand*\contentsname{Table of contents}
\else
  \newcommand\contentsname{Table of contents}
\fi
\ifdefined\listfigurename
  \renewcommand*\listfigurename{List of Figures}
\else
  \newcommand\listfigurename{List of Figures}
\fi
\ifdefined\listtablename
  \renewcommand*\listtablename{List of Tables}
\else
  \newcommand\listtablename{List of Tables}
\fi
\ifdefined\figurename
  \renewcommand*\figurename{Figure}
\else
  \newcommand\figurename{Figure}
\fi
\ifdefined\tablename
  \renewcommand*\tablename{Table}
\else
  \newcommand\tablename{Table}
\fi
}
\@ifpackageloaded{float}{}{\usepackage{float}}
\floatstyle{ruled}
\@ifundefined{c@chapter}{\newfloat{codelisting}{h}{lop}}{\newfloat{codelisting}{h}{lop}[chapter]}
\floatname{codelisting}{Listing}

\makeatother
\makeatletter
\@ifpackageloaded{caption}{}{\usepackage{caption}}
\@ifpackageloaded{subcaption}{}{\usepackage{subcaption}}
\makeatother

\ifLuaTeX
  \usepackage{selnolig}  
\fi
\usepackage[]{natbib}
\usepackage{bookmark}
\usepackage{multirow}
\def\pr{\textsf{P}}                    
\def\ep{\textsf{E}}
\def\var{\textsf{Var}}       
\def\cov{\textsf{Cov}}

\usepackage{amsthm}
\usepackage{algorithm}
\usepackage{algorithmic}
\usepackage{tabularx}
\newtheorem{theorem}{Theorem}

\newtheorem{lemma}{Lemma}
\newtheorem{condition}{Condition}

\newtheorem{remark}{Remark}
\newtheorem{assumption}{Assumption}

\IfFileExists{xurl.sty}{\usepackage{xurl}}{} 
\hypersetup{
  pdftitle={Title},
  pdfauthor={Author 1; Author 2},
  pdfkeywords={3 to 6 keywords, that do not appear in the title},
  colorlinks=true,
  linkcolor={blue},
  filecolor={Maroon},
  citecolor={Blue},
  urlcolor={Blue},
  pdfcreator={LaTeX via pandoc}}

\allowdisplaybreaks

\newcommand{\anon}{1}

\begin{document}

\def\spacingset#1{\renewcommand{\baselinestretch}%
{#1}\small\normalsize} \spacingset{1}


\if1\anon
{
  \title{\bf Change-Point Detection With Multivariate Repeated Measures}
  \author{Serim Han\\
    Department of Industrial and Systems Engineering, KAIST\\
    and \\
    Jingru Zhang \\
    School of Data Science, Fudan University\\
    and \\
    Hoseung Song \\
    Department of Industrial and Systems Engineering, KAIST\\}
  \maketitle
} \fi

\if0\anon
{
  \bigskip
  \bigskip
  \bigskip
  \begin{center}
    {\LARGE\bf Change-Point Detection With Multivariate Repeated Measures}
\end{center}
  \medskip
} \fi

\bigskip
\begin{abstract}
Graph-based methods have shown particular strengths in change-point detection (CPD) tasks for high-dimensional nonparametric settings. However, existing CPD research has rarely addressed data with repeated measurements or local group structures. A common treatment is to average repeated measurements, which can result in the loss of important within-individual information. In this paper, we propose a new graph-based method for detecting change-points in data with repeated measurements or local structures by incorporating both within-individual and between-individual information. Analytical approximations to the significance of the proposed statistics are derived, enabling efficient computation of $p$-values for the combined test statistic. We also establish consistency of the proposed test and the estimated change-point location. The proposed method effectively detects change-points across a wide range of alternatives, particularly when within-individual differences are present. The new method is illustrated through an analysis of the New York City taxi dataset.
\end{abstract}

\noindent%
\vfill

\newpage
\spacingset{1.8} 


\section{Introduction}

Change-point detection (CPD) is a fundamental problem in sequential data analysis, where the goal is to identify points in the sequence where the underlying probability distribution changes. CPD has diverse applications in various fields, including healthcare, environmental science, and image or text analysis (\citealp{aminikhanghahi2017survey}). Parametric approaches have been widely employed for univariate and low-dimensional data, as they provide effective detection under relatively simple distributional assumptions (see \citealp{carlstein1994change, chen2000parametric}). However, such parametric constraints often limit their applicability to real-world data, particularly in high-dimensional multivariate settings.

To address these limitations, a few nonparametric methods have been proposed. Rank-based methods (\citealp{pettitt1979non, lombard1987rank, ross2012two, zhou2025asymptotic}) rely on ordering information to detect distributional shifts. Kernel-based methods (\citealp{desobry2005online, harchaoui2008kernel, li2015m, song2024practical}) capture complex changes by projecting data into a feature space. Inter-point distance-based methods (\citealp{matteson2014nonparametric, li2020asymptotic}) utilize pairwise distances to detect changes. Graph-based methods (\citealp{chen2015graph, shi2017consistent, chu2019asymptotic, song2022asymptotic}) construct similarity graphs and test for structural changes. Beyond these approaches, other works have addressed related challenges: \cite{Saatcci2010gaussian} introduced Gaussian process change-point models that provide a flexible Bayesian nonparametric framework; \cite{wang2013locality} developed locality-based test statistics for identifying change-points in graph time series; and \cite{dubey2020frechet} proposed a computationally efficient Fréchet change-point method for metric-space-valued data. More recently, machine learning-based approaches have been increasingly developed, with advances fostered by deep learning methods. \cite{jin2019one} proposed a one-class SVM calibration method, and \cite{londschien2023random} developed a random forest framework with theoretical guarantees. Deep neural network-based methods (\citealp{ahmadzadeh2018change, li2024automatic}) leverage neural architectures to automatically learn representations for effective detection in high-dimensional data.

In many real-world applications, sequential data for CPD involves repeated measurements. For example, in healthcare or medical applications, physical activity data is often collected continuously over multiple days for one person. In such settings, for each time point $i = 1, \ldots, n$ with $\ell$ repeated measures, the observation can be represented as an $\ell$-dimensional vector $\mathbf{y}_i = (y_{i1}, y_{i2}, \ldots, y_{i\ell})^\top$. A common approach is to take the average of the repeated measures $\bar{y}_i = ({1}/{\ell}) \sum_{j=1}^{\ell} y_{ij}$, and treat it as a single value for individual $i$. Another approach is to treat all $\ell$ measurements as independent features. However, both approaches can lead to a loss of important information, and standard CPD methods may fail to detect meaningful signals arising from the variability among repeated measurements.

\begin{figure}[htbp]
    \centering
    \includegraphics[width=0.8\textwidth]{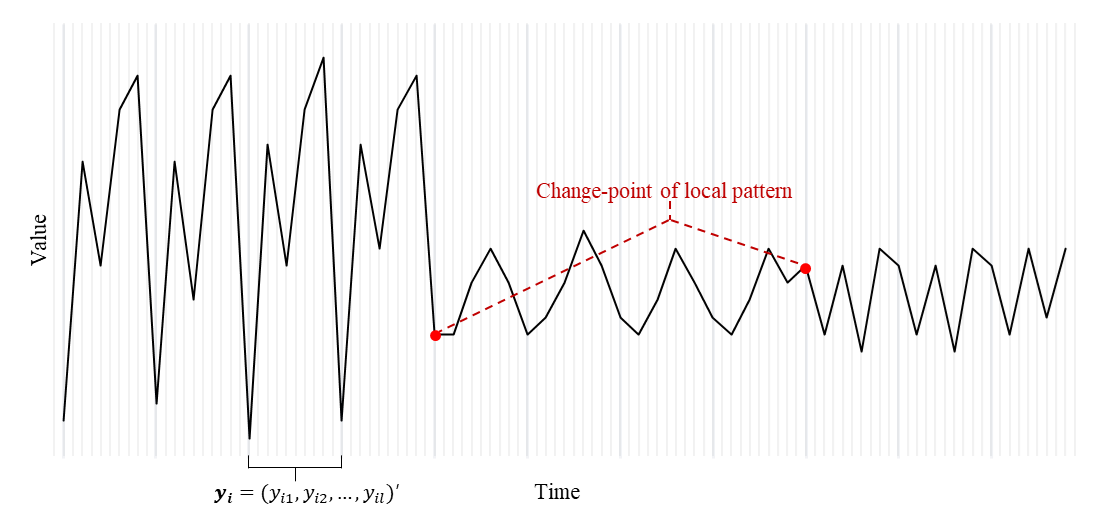}
    \caption{An example of a time series with changes in local structure. Vertical stripes denote time points, the black curve shows observations, and red markers indicate local change-points.}
    \label{fig:local}
\end{figure}

Repeated measurements can be viewed more generally as local group structures, where similar individuals or temporally adjacent observations are strongly correlated. In longitudinal or panel studies, local patterns may recur across individuals observed at the same time. Similarly, time series may exhibit short-term periodic patterns such as weekly or monthly cycles. Figure~\ref{fig:local} illustrates a setting with a change in local structure. Although the global trend is relatively stable, local structural changes occur at specific time points. A common practice is to aggregate local patterns, but this may obscure important within-group information. Several methods have been proposed for repeated measurements or local patterns. \cite{lee1991sequential} introduced a group sequential test based on a linear mixed-effects model for comparing rates of change between two treatment groups, while \cite{horvath2012change} proposed a CUSUM-based test for mean shifts in high-dimensional panel data. These methods are parametric and rely on strong assumptions, limiting their applicability to complex or high-dimensional data.

Recently, \cite{zhang2022two} proposed a graph-based segmentation method for repeated measurements, but it does not account for sequential structure. If a sequence contains strong local dependence, their framework can be extended to CPD methods that explicitly model within-group structure and detect changes in local patterns.


\subsection{Our Contributions}
We propose a novel graph-based method for change-point detection in data with repeated measurements or locally structured sequences. Building upon the framework of graph-based statistics with repeated measures introduced by \cite{zhang2022two}, we extend the two-sample test setting to the change-point detection scenario. Specifically, we construct a max-type combination of within-individual and between-individual statistics, enabling the new test to detect change-points driven by either type of structural change. We also establish the limiting distributions of the proposed stochastic processes and derive analytical $p$-value approximations with the implementation of skewness correction, facilitating application to large datasets. In addition, we show that the proposed test is consistent and that the resulting change-point estimator converges to the true change-point location. The codes for reproducing simulation results are available at \url{https://github.com/haser66/gSeg_repeated}.

The organization of the paper is as follows. In Section~\ref{sec:newstat}, we propose three new scan statistics to accommodate repeated measurements and local group structures in the sequence. The asymptotic behavior of the new scan statistics, the analytical $p$-value approximations, a proposed test and its consistency are presented in Section~\ref{sec:asym}. Simulation results through extensive experiments are presented in Section~\ref{sec:simul}, and in Section~\ref{sec:nyc} we analyze the New York City taxi dataset. We conclude with a discussion in Section \ref{sec:disc}.



\section{New Scan Statistics} \label{sec:newstat}

We consider a sequence of independent observations
\begin{equation*}
\mathbf{y}_i = (y_{i1}, y_{i2}, \ldots, y_{i\ell})^T, \quad i = 1, \ldots, n,
\end{equation*}
where each $y_{ij} \in \mathbb{R}^d$, $j=1,\ldots,\ell$, represents a $d$-dimensional measurement taken at the $j$-th repeated measure, and $\ell$ denotes the number of repeated measures for each individual. Thus, $\mathbf{y}_i$ can be viewed as a sequence of $\ell$ vectors in $\mathbb{R}^d$, equivalently represented as an $\ell \times d$ matrix. We aim to test the null hypothesis
\begin{equation} \label{hypo}
H_0 : \mathbf{y}_i \sim F_0, \quad i = 1, \ldots, n,
\end{equation}
against the single change-point alternative
\begin{equation}
H_1 : \exists\ 1 \leq \tau < n, \quad
\mathbf{y}_i \sim
\begin{cases}
F_0 & \text{if } i \leq \tau, \\
F_1 & \text{otherwise}.
\end{cases}
\end{equation}

To preserve the structure of repeated measurements in the graph-based approach, we construct a similarity graph $G$, such as a minimum spanning tree\footnote{An MST is a spanning tree that connects all observations minimizing the sum of the distances of edges.} (MST) or nearest neighbor graph (NNG), based on all repeated measurements across individuals. In particular, we adopt the $k$-MST\footnote{A $k$-MST is the union of the first through the $k$th MSTs. The $1$st MST is the usual minimum spanning tree, and each $j$th MST ($j>1$) connects all observations and minimizes the sum of the distances of all edges without using any edge from the previous $1$st MST, $\cdots$, and $(j-1)$th MST.} algorithm to construct the graph on $n \times \ell$ nodes, representing $n$ individuals with $\ell$ repeated measures per individual. Following the recommendation of \cite{chen2018weighted}, we set $k = 9$, thereby using the 9-MST as the similarity graph. We then divide $G$ into two disjoint subgraphs: between-individual graph $G_{\text{out}}$ and within-individual graph $G_{\text{in}}$. $G_{\text{out}}$ contains edges connecting nodes from different individuals and $G_{\text{in}}$ contains edges connecting nodes belonging to the same individual.

\begin{figure}[h]
    \centering
    \includegraphics[width=0.8\textwidth]{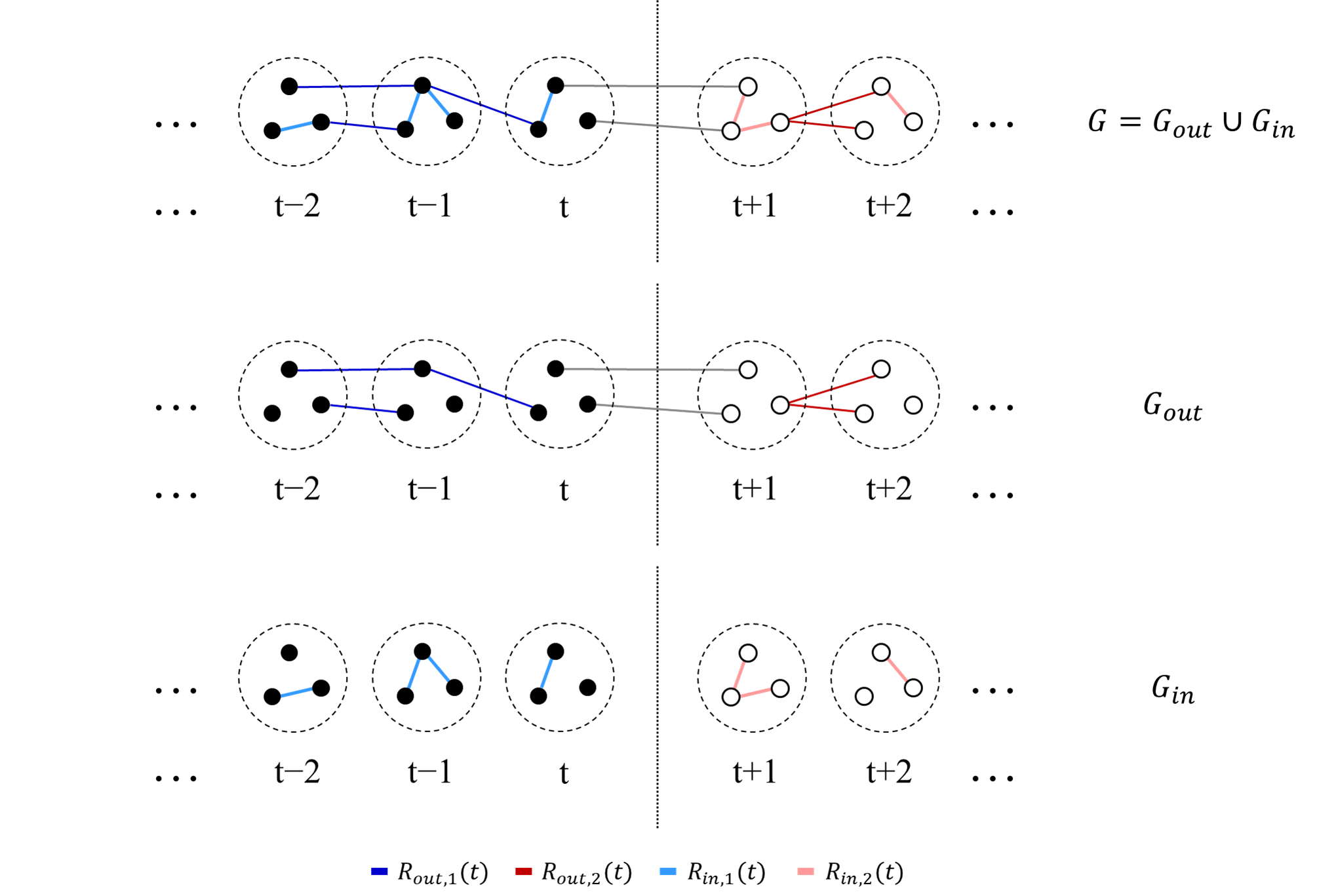}
    \caption{Examples of $G, G_{out},$ and $ G_{in}$ with $\ell=3$.}
    \label{fig:graph}
\end{figure}

We define the indicator function for any node $i$ observed after a time point $t$ as $g_i(t) = I_{i > t}$. For an edge $e = (i,j) \in G$, the edge type $J_e(t)$ is given by
\begin{equation*}
J_e(t) =
\begin{cases}
0, & \text{if } g_i(t) \neq g_j(t), \\
1, & \text{if } g_i(t) = g_j(t) = 0, \\
2, & \text{if } g_i(t) = g_j(t) = 1.
\end{cases}
\end{equation*}
For any candidate value $t$ of $\tau$, we define
\begin{equation}
R_{\text{out}, k}(t) = \sum_{e \in G_\text{out}} I_{J_e(t) = k}, \quad R_{\text{in}, k}(t) = \sum_{e \in G_\text{in}} I_{J_e(t) = k}, \quad k = 1, 2.
\end{equation}
Accordingly, $R_{\text{out}, 1}(t)$ and $R_{\text{in}, 1}(t)$ represent the numbers of between-individual edges and within-individual edges connecting observations prior to $t$, respectively. Similarly, $R_{\text{out}, 2}(t)$ and $R_{\text{in}, 2}(t)$ denote the corresponding numbers of edges connecting observations after $t$. Figure~\ref{fig:graph} illustrates the decomposition of $G$ into $G_{\text{out}}$ and $G_{\text{in}}$. Edges connecting nodes prior to $t$ are depicted in blue tones, while those connecting nodes after $t$ are red-toned. Dark blue and dark red lines represent the edges between different individuals and thus belong to $G_{\text{out}}$. Light blue and light red lines indicate edges within the same individual and are included in $G_{\text{in}}$. Gray lines correspond to the edges connecting the nodes before and after $t$.

We evaluate the extent to which the observed test statistic deviates from its permutation null distribution. Under $H_0$, the $n$ individuals are permuted, with each permutation having equal probability $1/n!$. In this procedure, the repeated measurements of the same individual are treated as a single unit and are permuted jointly. In other words, if an individual $\mathbf{y}_i$ is assigned to a particular group in a given permutation, all of its repeated measurements are simultaneously assigned to that group. Unless otherwise specified, we use $\pr$, $\ep$, $\var$, and $\cov$ to denote probability, expectation, variance, and covariance under the permutation null distribution, respectively.

Let $|G|$, $|G_{\text{out}}|$, and $|G_{\text{in}}|$ be the number of edges in graphs $G$, $G_{\text{out}}$, and $G_{\text{in}}$, respectively. As illustrated in Figure~\ref{fig:graph}, the set $G_{\text{out}}$ consists of three types of edges: dark blue edges ($R_{\text{out},1}(t)$), dark red edges ($R_{\text{out},2}(t)$), and the remaining gray edges. Since the number of gray edges is determined by $|G_{\text{out}}|-R_{\text{out},1}(t)-R_{\text{out},2}(t)$, the pair $(R_{\text{out},1}(t), R_{\text{out},2}(t))$ fully characterizes the information in $G_{\text{out}}$. Similarly, the set $G_{\text{in}}$ consists of light blue edges ($R_{\text{in},1}(t)$) and light red edges ($R_{\text{in},2}(t)$). Since $R_{\text{in},2}(t)$ is determined by $|G_{\text{in}}|-R_{\text{in},1}(t)$, it is sufficient to use $R_{\text{in},1}(t)$ alone to capture all the information contained in $G_{\text{in}}$. 

To capture changes in between-individual connectivity, we consider two test statistics $R_{\text{out},w}(t)$ and $R_{\text{out},d}(t)$, which are linear combinations of $R_{\text{out},1}(t)$ and $R_{\text{out},2}(t)$, 
\begin{align*}
R_{\text{out},w}(t) &= \frac{(n-t-1) R_{\text{out},1}(t) + (t - 1) R_{\text{out},2}(t)}{n-2}, \\
R_{\text{out},d}(t) &= R_{\text{out},1}(t) - R_{\text{out},2}(t).
\end{align*}
We would expect $R_{\text{out},w}(t)$ to detect location changes, whereas $R_{\text{out},d}(t)$ would be sensitive to scale changes. Taking into account changes in within-individual connectivity as well, we employ $R_{\text{out},w}(t), R_{\text{out},d}(t),$ and $R_{\text{in},1}(t)$ to comprehensively capture the change-point signal in repeated measurements. The corresponding standardized versions of these statistics are
\begin{align*}
&Z_{\text{out},w}(t) = \frac{R_{\text{out},w}(t) - \mu_{\text{out},w}(t)}{\sigma_{\text{out},w}(t)}, \
Z_{\text{out},d}(t) = \frac{R_{\text{out},d}(t) - \mu_{\text{out},d}(t)}{\sigma_{\text{out},d}(t)}, \ Z_{\text{in}}(t) =  \frac{R_{\text{in},1}(t) - \mu_{\text{in}}(t)}{\sigma_{\text{in}}(t)},
\end{align*}
where
\begin{align*}
&\mu_{\text{out},w}(t) = \ep(R_{\text{out},w}(t)), &&\mu_{\text{out},d}(t) = \ep(R_{\text{out},d}(t)), &&\mu_{\text{in}}(t) = \ep(R_{\text{in},1}(t)), \\
&\sigma_{\text{out},w}(t) = \sqrt{\var(R_{\text{out},w}(t))}, &&\sigma_{\text{out},d}(t) = \sqrt{\var(R_{\text{out},d}(t))}, &&\sigma_{\text{in}}(t) = \sqrt{\var(R_{\text{in},1}(t))}.
\end{align*}
Under the alternative hypothesis, $Z_{\text{out},w}(t)$ tends to be large for location shifts in between-individuals, while $|Z_{\text{out},d}(t)|$ tends to be large for scale changes in between-individuals. $|Z_{\text{in}}(t)|$ becomes large for changes that occur within individuals. Considering these three statistics together, a wide range of distributional changes can be effectively detected.

We define a symmetric matrix $D = (D_{uv})_{n \times n}$, where $D_{uv}$ is the number of edges between individuals $u$ and $v$ in $G$. For each individual $u$, we denote by $D_u = \sum_{v \neq u} D_{uv}$ the total number of edges that connect individual $u$ to others. The following lemma provides analytic expressions for $\ep$, $\var$, and $\cov$ without explicitly performing permutations.

\begin{lemma}\label{lemma:evar} Under the permutation null distribution, we have
\begin{align*}
&\ep(R_{\text{out},1}(t)) = \frac{t(t-1)}{n(n-1)} |G_{\text{out}}| , \quad
\ep(R_{\text{out},2}(t)) = \frac{(n-t)(n-t-1)}{n(n-1)} |G_{\text{out}}|, \\
&\ep(R_{\text{in},1}(t)) = \frac{t}{n} |G_{\text{in}}|, \\
&\var(R_{\text{out},1}(t)) = \frac{t(t - 1)(n-t)(n-t - 1)}{n(n - 1)(n - 2)(n - 3)} \\
& \quad \qquad \qquad \qquad \times \left\{ \sum_{u \neq v} \frac{D_{uv}^2}{2} 
+ \frac{t-2}{n-t-1} \left( \sum_u D_u^2 - \frac{4}{n} |G_{\text{out}}|^2 \right)
- \frac{2}{n(n - 1)} |G_{\text{out}}|^2 \right\}, \\
&\var(R_{\text{out},2}(t)) = \frac{t(t - 1)(n-t)(n-t - 1)}{n(n - 1)(n - 2)(n - 3)} \\
& \quad \qquad \qquad \qquad \times \left\{ \sum_{u \neq v} \frac{D_{uv}^2}{2} 
+ \frac{n-t-2}{t-1} \left( \sum_u D_u^2 - \frac{4}{n} |G_{\text{out}}|^2 \right)
- \frac{2}{n(n - 1)} |G_{\text{out}}|^2 \right\}, \\
&\var(R_{\text{in},1}(t)) = \frac{t(n-t)}{n(n - 1)} \left( \sum_u D_{uu}^2 - \frac{|G_{\text{in}}|^2}{n} \right), \\
&\cov(R_{\text{out},1}(t), R_{\text{out},2}(t)) = 
\frac{t(t - 1)(n-t)(n-t - 1)}{n(n - 1)(n - 2)(n - 3)} \\
& \quad \qquad \qquad \qquad \qquad \qquad  \times \left\{ \sum_{u \neq v} \frac{D_{uv}^2}{2} 
- \left( \sum_u D_u^2 - \frac{4}{n} |G_{\text{out}}|^2 \right) 
- \frac{2}{n(n - 1)} |G_{\text{out}}|^2 \right\}, \\
&\cov(R_{\text{out},1}(t), R_{\text{in},1}(t)) = 
\frac{t(t - 1)(n-t)}{n(n - 1)(n - 2)} 
\left( \sum_{u=1}^n D_{uu} D_u - \frac{2}{n} |G_{\text{in}}| |G_{\text{out}}| \right), \\
&\cov(R_{\text{out},2}(t), R_{\text{in},1}(t)) =
- \frac{t(n-t)(n-t-1)}{n(n - 1)(n - 2)} 
\left( \sum_{u=1}^n D_{uu} D_u - \frac{2}{n} |G_{\text{in}}| |G_{\text{out}}| \right).
\end{align*}
\end{lemma}
The proof is provided in Section A.1 in the supplementary materials.

Based on Lemma~\ref{lemma:evar}, each statistic is standardized such that its expectation equals zero and variance equals one. The covariances are given by $\cov(Z_{\text{out},w}(t), Z_{\text{out},d}(t)) = 0$ and $\cov(Z_{\text{out},w}(t), Z_{\text{in}}(t))=0.$ Note that the covariance between $Z_{\text{out},d}(t)$ and $Z_{\text{in}}(t)$ is not generally zero; we will address this correlation in Section \ref{sec:ortho}.


\section{Asymptotics of the Scan Statistics and the New Test} \label{sec:asym}
To conduct change-point testing, we evaluate the significance of the proposed scan statistics. We combine $Z_{\text{out},w}(t), |Z_{\text{out},d}(t)|, \text{ and } |Z_{\text{in}}(t)|$ into a unified scan statistic and propose a new test based on its tail probability. While permutation procedures are feasible for small samples, they can be time-consuming for large $n$. Hence, we develop analytic expressions for fast and practical implementation. Specifically, we derive the asymptotic distribution of the scan statistics and improve the accuracy through skewness correction.


\subsection{Asymptotic Null Distributions of the Basic Processes} \label{subsection:null}

In this section, we first derive the limiting distributions of $\{ Z_{\text{out},w}([nu]) : 0<u<1 \}$, $\{ Z_{\text{out},d}([nu]) : 0<u<1 \}$, and $\{ Z_{\text{in}}([nu]) : 0<u<1 \}$. Here, $[x]$ denotes the largest integer that is no larger than $x$. Before proceeding, we specify the notation and assumptions of the setting. Let $C_u$ denote the set of repeated measurements from individual $u$. For an edge $e = (i,j) \in G_{\text{out}}$, where $i \in C_u$, $j \in C_v$, and $u \neq v$, let
\begin{align*}
A_{\text{out},e} 
= \{e\} 
&\cup \{e' = (k, \ell) \in G_{\text{out}} \mid k \in C_u \cup C_v \text{ or } \ell \in C_u \cup C_v\} \\
&\cup \{e'' = (k, \ell) \in G_{\text{in}} \mid k, \ell \in C_u \cup C_v\},
\end{align*}
and for an edge $ e = (i,j) \in G_{\text{in}}$ with $ \ i,j \in C_u$, define
\begin{equation*}
A_{\text{in},e} 
= \{e' = (k, \ell) \in G_{\text{out}} \mid k \in C_u \text{ or } \ell \in C_u\} 
\cup \{e'' = (\ell, k) \in G_{\text{in}} \mid k, \ell \in C_u\}.
\end{equation*}
We then set
\begin{equation*}
A_e = 
\begin{cases}
A_{\text{out},e}, & \text{if } e \in G_{\text{out}}, \\
A_{\text{in},e}, & \text{if } e \in G_{\text{in}}.
\end{cases}
\end{equation*}
The corresponding set $B_e$ is defined in a similar way:
\begin{equation*}
B_{\text{out},e} = \bigcup_{\tilde{e} \in A_{\text{out},e}} A_{\tilde{e}}, \qquad
B_{\text{in},e} = \bigcup_{\tilde{e} \in A_{\text{in},e}} A_{\tilde{e}}, \qquad
B_e = 
\begin{cases}
B_{\text{out},e}, & \text{if } e \in G_{\text{out}}, \\
B_{\text{in},e}, & \text{if } e \in G_{\text{in}}.
\end{cases}
\end{equation*}

We assume $t = O(n)$, $n - t = O(n)$ and $\ell = O(1)$. Throughout, we write $a_n = O(b_n)$ when $a_n$ and $b_n$ are of the same order, and write $a_n = o(b_n)$ when $a_n$ is of smaller order than $b_n$. We proceed to introduce the following conditions.

\begin{condition} \label{con:1} $|G_{\text{out}}| = O(n), \ |G_{\text{in}}| = O(n).$
\end{condition}
\begin{condition} \label{con:2} $\sum_u D_u^2 - \frac{4|G_{\text{out}}|^2}{n}=O(n), \ \sum_u D_{uu}^2 - \frac{|G_{\text{in}}|^2}{n} = O(n).$ \end{condition}
\begin{condition} \label{con:3} $\sum_{e \in G_{\text{out}}} |A_{\text{out},e}|\,|B_{\text{out},e}| = o(n^{1.5}).$
\end{condition}

\begin{remark}\label{remark:cond} Condition~\ref{con:1} requires that the numbers of edges in $G_{\text{out}}$ and $G_{\text{in}}$ grow on the same order as $n$ and can be easily satisfied with an appropriate choice of $G$. Condition~\ref{con:2} ensures that $(R_{\text{out},1}(t), R_{\text{out},2}(t), R_{\text{in},1}(t))^\top$ does not degenerate asymptotically. For example, since 
\begin{equation*}
\sum_u D_u^2 - \frac{4|G_{\text{out}}|^2}{n} = \sum_u \left( D_u - \frac{2|G_{\text{out}}|}{n} \right)^2, 
\quad
\sum_u D_{uu}^2 - \frac{|G_{\text{in}}|^2}{n} = \sum_u \left( D_{uu} - \frac{|G_{\text{in}}|}{n} \right)^2,
\end{equation*}
if $D_u - 2|G_{\text{out}}|/n = O(1)$ and $D_{uu} - |G_{\text{in}}|/n = O(1)$, Condition~\ref{con:2} is satisfied. Condition~\ref{con:3} requires that the graph has no large hub or a cluster of small hubs, where a hub is a node with a large degree. Similar conditions appears in the context of graph-based statistics for independent observations \citep{chen2017new, chen2018weighted}.
\end{remark}

\begin{remark}\label{remark:in}
The condition $|G_{\text{in}}| = O(n)$ requires that repeated measurements contain sufficient common information, ensuring that within-individual edges grow at a meaningful rate and can be explored by our test statistic $Z_{\text{in}}(t)$. We also assume $\ell = O(1)$, i.e., as $n\to\infty$, the number of repeated measurements per individual remains bounded. This prevents a large hub from dominating the majority of edges in $G$ and ensures that the regularity conditions, such as Condition~\ref{con:3}, hold. 
\end{remark}

Now we denote some limits. Conditions~\ref{con:1} and~\ref{con:2} imply that $\sum_u D_u^2 = O(n)$ and $\sum_u D_{uu}^2 = O(n)$. In addition, since 
\begin{equation*}
2|G_{\text{out}}| = \sum_{u \ne v} D_{uv} \le \sum_{u \ne v} D_{uv}^2 \le \sum_u D_u^2,
\quad
|G_{\text{in}}| = \sum_u D_{uu} \le \sum_u D_{uu} D_u \le \sqrt{\sum_u D_{uu}^2 \sum_u D_u^2},
\end{equation*}
it follows that $\sum_{u \ne v} D_{uv}^2 = O(n)$ and $\sum_u D_{uu} D_u = O(n).$

\begin{assumption} \label{ass:limit}
We assume that the following limits exist as $n \to \infty$:
\begin{align*}
    \lim_{n \to \infty} \frac{|G_{\text{out}}|}{n} &= b_1, &
    \lim_{n \to \infty} \left( \frac{\sum_u D_u^2}{n} - \frac{4|G_{\text{out}}|^2}{n^2} \right) &= b_2, &
    \lim_{n \to \infty} \frac{\sum_{u \neq v} D_{uv}^2}{n} &= b_3, \\
    \lim_{n \to \infty} \frac{|G_{\text{in}}|}{n} &= b_4, &
    \lim_{n \to \infty} \left( \frac{\sum_u D_{uu}^2}{n} - \frac{|G_{\text{in}}|^2}{n^2} \right) &= b_5, &
    \lim_{n \to \infty} \frac{\sum_u D_{uu} D_u}{n} &= b_6.
\end{align*}
\end{assumption}

\begin{theorem}\label{thm:gauss} Under Conditions~\ref{con:1}--\ref{con:3} and Assumption~\ref{ass:limit}, as $n \to \infty$,
\begin{enumerate}
    \item $ \{ Z_{\text{out},w}([nu]) : 0<u<1 \}$ and $\{Z_{\text{out},d}([nu]): 0<u<1 \}$ converge to independent Gaussian processes in finite dimensional distributions, which we denote as $ \{ Z_{\text{out},w}^*(u) : 0<u<1 \}$ and $\{Z_{\text{out},d}^*(u): 0<u<1 \}$, respectively.
    \item $ \{ Z_{\text{out},w}([nu]) : 0<u<1 \}$ and $\{Z_{\text{in}}([nu]): 0<u<1 \}$ converge to independent Gaussian processes in finite dimensional distributions, denoted by $ \{ Z_{\text{out},w}^*(u) : 0<u<1 \}$ and $\{Z_{\text{in}}^*(u): 0<u<1 \}$, respectively.

\end{enumerate}
\end{theorem}
The proof is provided in Section A.2 in the supplementary materials.

Let $\rho_{\text{out},w}^*(u, v) = \cov(Z_{\text{out},w}^*(u), Z_{\text{out},w}^*(v))$, $\rho_{\text{out},d}^*(u, v) = \cov(Z_{\text{out},d}^*(u), Z_{\text{out},d}^*(v))$ and $\rho_{\text{in}}^*(u, v) = \cov(Z_{\text{in}}^*(u), Z_{\text{in}}^*(v))$. The exact covariance functions of the limiting Gaussian processes $\{Z_{\text{out},w}^*(u):\ 0 < u < 1\}$, $\{Z_{\text{out},d}^*(u):\ 0 < u < 1\}$, and $\{Z_{\text{in}}^*(u):\ 0 < u < 1\}$ are stated in the following theorem.

\begin{theorem}\label{thm:rho} The exact expressions for $\rho_{\text{out},w}^*(u, v)$, $\rho_{\text{out},d}^*(u, v)$ and $\rho_{\text{in}}^*(u, v)$ are
\begin{equation*}
\rho_{\text{out},w}^*(u, v) = \frac{(u \wedge v)(1 - (u \vee v))}{(u \vee v)(1 - (u \wedge v))},
\end{equation*}
\begin{equation*}
\rho_{\text{out},d}^*(u, v) = \rho_{\text{in}}^*(u, v) = \frac{(u \wedge v)(1 - (u \vee v))}{
\sqrt{(u \wedge v)(1 - (u \wedge v))(u \vee v)(1 - (u \vee v))}},
\end{equation*}
\textit{where $u \wedge v = \min(u, v)$ and $u \vee v = \max(u, v)$.}
\end{theorem}
The proof of Theorem~\ref{thm:rho} can be found in Section A.3 in the supplementary materials.


\subsection{Analytical $P$-value Approximation} \label{subsection:pval}

Based on Theorems~\ref{thm:gauss} and~\ref{thm:rho}, we approximate the tail probabilities through the Woodroofe’s method (\citealp{woodroofe1976frequentist}, \citeyear{woodroofe1978large}) and Siegmund’s method (\citealp{siegmund1988approximate}, \citeyear{siegmund1992tail}). Specifically, following similar arguments in the proof for Proposition 3.4 in \cite{chen2015graph}, when Conditions~\ref{con:1}--\ref{con:3} hold, $n, b, n_0, n_1 \to \infty$ in a way such that for some $b_0 > 0$ and $0 < x_0 < x_1 < 1$, $b/\sqrt{n} \to b_0$, $n_0/n \to x_0$ and $n_1/n \to x_1$, as $n \to \infty$, we have
\begin{align*}
\pr\left( \max_{n_0 \le t \le n_1} Z_{\text{out},w}^*(t/n) > b \right)
&\sim b\phi(b) \int_{x_0}^{x_1} h_{\text{out},w}^*(x) \ \nu \left( b_0 \sqrt{2h_{\text{out},w}^*(x)} \right) \, dx, \\
\pr\left( \max_{n_0 \le t \le n_1} \left| Z_{\text{out},d}^*(t/n) \right| > b \right)
&\sim 2b\phi(b) \int_{x_0}^{x_1} h_{\text{out},d}^*(x) \ \nu \left(b_0 \sqrt{2h_{\text{out},d}^*(x)} \right) \, dx, \\
\pr\left( \max_{n_0 \le t \le n_1} \left| Z_{\text{in}}^*(t/n) \right| > b \right)
&\sim 2b\phi(b) \int_{x_0}^{x_1} h_{\text{in}}^*(x) \ \nu \left(b_0 \sqrt{2h_{\text{in}}^*(x)} \right) \, dx,
\end{align*}
where
\begin{align*}
h_{\text{out},w}^*(x) &= \lim_{u \nearrow x} \frac{\partial \rho_{\text{out},w}^*(u,x)}{\partial u} 
= -\lim_{u \searrow x} \frac{\partial \rho_{\text{out},w}^*(u,x)}{\partial u},\\
h_{\text{out},d}^*(x) &= \lim_{u \nearrow x} \frac{\partial \rho_{\text{out},d}^*(u,x)}{\partial u}
= - \lim_{u \searrow x} \frac{\partial \rho_{\text{out},d}^*(u,x)}{\partial u},\\
h_{\text{in}}^*(x) &= \lim_{u \nearrow x} \frac{\partial \rho_{\text{in}}^*(u,x)}{\partial u}
= - \lim_{u \searrow x} \frac{\partial \rho_{\text{in}}^*(u,x)}{\partial u}.
\end{align*}
Here, $\nu(x)$ can be numerically approximated as
\begin{equation*}
\nu(x) \approx \frac{(2/x)(\Phi(x/2) - 0.5)}{(x/2)\Phi(x/2) + \phi(x/2)}
\end{equation*}
by \cite{siegmund2007statistics}, where $\Phi(\cdot)$ and $\phi(\cdot)$ are the standard normal cumulative density function and standard normal density function, respectively.
It can be shown that
\begin{equation*}
h_{\text{out}, w}^*(x) = \frac{1}{x(1 - x)}, \quad
h_{\text{out}, d}^*(x) = h_{\text{in}}^*(x) =  \frac{1}{2x(1 - x)}.
\end{equation*}
Based on the above results, we can approximate the tail probabilities by
\begin{align}
\pr\left( \max_{n_0 \le t \le n_1} Z_{\text{out},w}(t) > b \right) &\approx b\phi(b) \int_{x_0}^{x_1} h_{\text{out},w}^*(x) \ \nu \left( b \sqrt{2h_{\text{out},w}^*(x)/n} \right) \, dx, \label{eq:Pow} \\
\pr\left( \max_{n_0 \le t \le n_1} \left| Z_{\text{out},d}(t) \right| > b \right) &\approx 2b\phi(b) \int_{x_0}^{x_1} h_{\text{out},d}^*(x) \ \nu \left(b \sqrt{2h_{\text{out},d}^*(x)/n} \right) \, dx, \label{eq:Pod} \\
\pr\left( \max_{n_0 \le t \le n_1} \left| Z_{\text{in}}(t) \right| > b \right)
&\approx 2b\phi(b) \int_{x_0}^{x_1} h_{\text{in}}^*(x) \ \nu \left(b \sqrt{2h_{\text{in}}^*(x)/n} \right) \, dx. \label{eq:Pin}
\end{align}

\begin{remark}\label{remark:finite} In practice, when approximating the tail probabilities, we use $h_{\text{out},w}(n,x)$ in place of $h_{\text{out},w}^*(x)$, where $h_{\text{out},w}(n,x)$ is the finite-sample equivalent of $h_{\text{out},w}^*(x)$. That is,
\begin{equation*}
h_{\text{out},w}(n,x) = n \lim_{s \nearrow nx} \frac{\partial \rho_{\text{out},w}(s, nx)}{\partial s},
\end{equation*}
with $\rho_{\text{out},w}(s,t) := \cov(Z_{\text{out},w}(s), Z_{\text{out},w}(t))$. The explicit expression for $h_{\text{out},w}(n,x)$ can also be derived and simplified to be
\begin{equation*}
h_{\text{out},w}(n,x) = \frac{(n-1)(2nx^2 - 2nx + 1)}{2x(1 - x)(n^2x^2 - n^2x + n - 1)}.
\end{equation*}
It is clear from the above expression that $h_{\text{out},w}(n,x)$ does not depend on the graph $G$ as well. Also, it is easy to show that $\lim_{n \to \infty} h_{\text{out},w}(n,x) = h_{\text{out},w}^*(x)$.

The finite-sample equivalent version of $h_{\text{out},d}^*(x)$ has the same closed-form expression as $h_{\text{out},d}^*(x)$. This can be written as
\begin{equation*}
h_{\text{out},d}(n,x) = n \lim_{s \nearrow nx} \frac{\partial \cov(Z_{\text{out},d}(s), Z_{\text{out},d}([nx]))}{\partial s} = \frac{1}{2x(1 - x)}.
\end{equation*}
Similarly, $h_{\text{in}}(x)$ can be expressed in the same way as $h_{\text{out},d}(x)$. That is, $h_{\text{in}}(n,x) = {1}/{(2x(1 - x))}$.
\end{remark}


\subsection{Skewness Correction} \label{subsection:skew}

\begin{figure}[htbp]
    \centering
    \includegraphics[width=0.9\textwidth]{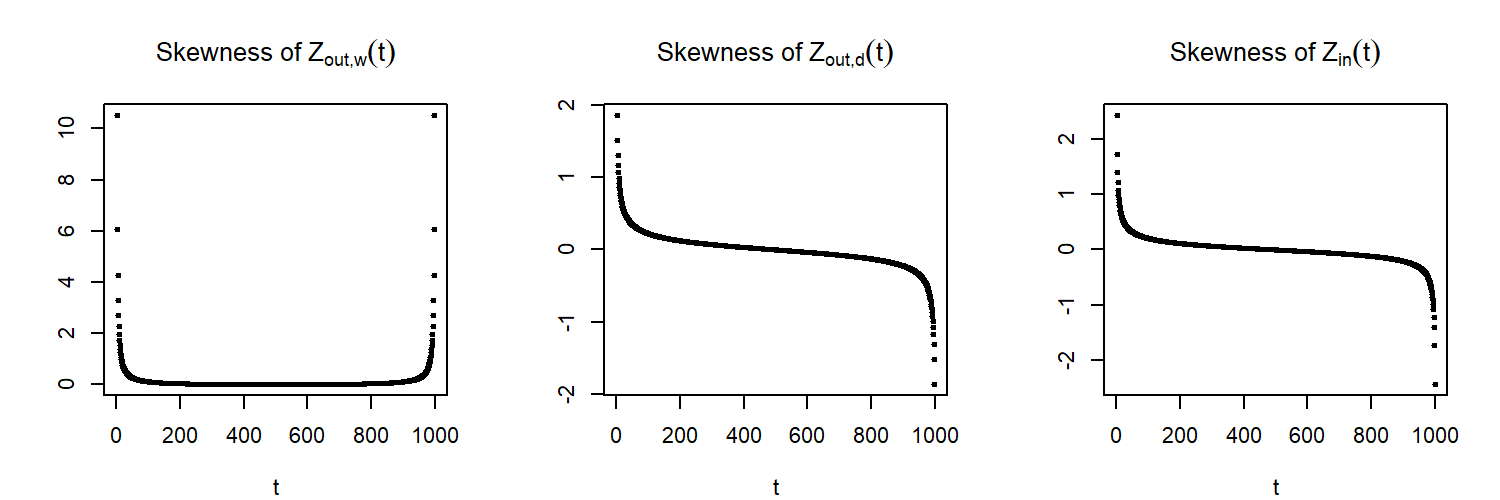}
    \caption{Plots of skewness $\gamma_{\text{out},w}(t)$, $\gamma_{\text{out},d}(t)$ and $\gamma_{\text{in}}(t)$ against $t$ with MST for a sequence of 1000 points randomly generated from the 50-dimensional multivariate Gaussian distribution with $\ell=5$.}
    \label{fig:skewness}
\end{figure}

As discussed in \cite{chen2015graph} and \cite{chu2019asymptotic}, the analytical approximations of the tail probabilities~(\ref{eq:Pow})--(\ref{eq:Pin}) can be improved by skewness correction as $n_0$ and $n_1$ are close to the two ends, particularly in high-dimensional settings or when the similarity graph contains hubs. This arises because, when $t/n$ is close to 0 or 1, $Z_{\text{out},w}(t)$, $Z_{\text{out},d}(t)$, and $Z_{\text{in}}(t)$ slowly converge to Gaussian processes. Figure~\ref{fig:skewness} displays the skewness of these statistics for a sequence of 1000 individuals, each with 5 repeated measures, generated from the multivariate Gaussian distribution. It can be clearly seen that $Z_{\text{out},w}(t)$ is strongly right-skewed near both ends of the sequence, whereas $Z_{\text{out},d}(t)$ and $Z_{\text{in}}(t)$ are right-skewed for small $t$ and left-skewed for large $t$.

To address this issue, since the skewness depends on the value of $t$, we employ the procedure described in~\cite{chen2015graph}. Let $\gamma_{\text{out},w}(t) = \ep(Z_{\text{out},w}^3(t))$, $\gamma_{\text{out},d}(t) = \ep(Z_{\text{out},d}^3(t))$, and $\gamma_{\text{in}}(t) = \ep(Z_{\text{in}}^3(t))$. The analytical $p$-value approximations after skewness correction are
\begin{align}
\pr\left( \max_{n_0 \le t \le n_1} Z_{\text{out},w}(t) > b \right)
&\approx b \phi(b) \int_{n_0/n}^{n_1/n} K_{\text{out},w}(nx) h_{\text{out},w}(n,x) \nu\left(b \sqrt{2 h_{\text{out},w}(n,x)/n}\right) dx, \label{eq:skew_Pow} \\
\pr\left( \max_{n_0 \le t \le n_1} |Z_{\text{out},d}(t)| > b \right)
&\approx 2b \phi(b) \int_{n_0/n}^{n_1/n} K_{\text{out},d}(nx) h_{\text{out},d}(n,x) \nu\left(b \sqrt{2 h_{\text{out},d}(n,x)/n}\right) dx, \label{eq:skew_Pod} \\
\pr\left( \max_{n_0 \le t \le n_1} |Z_{\text{in}}(t)| > b \right)
&\approx 2b \phi(b) \int_{n_0/n}^{n_1/n} K_{\text{in}}(nx) h_{\text{in}}(n,x) \nu\left(b \sqrt{2 h_{\text{in}}(n,x)/n}\right) dx, \label{eq:skew_Pin}
\end{align}
where
\begin{align*}
K_{\text{out},w}(t) &= \frac{\exp\left(\frac{1}{2}(b - \hat{\theta}_{b,w}(t))^2 + \frac{1}{6} \gamma_{\text{out},w}(t)\hat{\theta}_{b,w}(t)^3\right)}{\sqrt{1 + \gamma_{\text{out},w}(t)\hat{\theta}_{b,w}(t)}} \ \text{ with} \ \hat{\theta}_{b,w}(t) = \frac{-1 + \sqrt{1 + 2\gamma_{\text{out},w}(t)b}}{\gamma_{\text{out},w}(t)},\\
K_{\text{out},d}(t) &= \frac{\exp\left(\frac{1}{2}(b - \hat{\theta}_{b,d}(t))^2 + \frac{1}{6} \gamma_{\text{out},d}(t)\hat{\theta}_{b,d}(t)^3\right)}{\sqrt{1 + \gamma_{\text{out},d}(t)\hat{\theta}_{b,d}(t)}} \ \text{ with} \ \hat{\theta}_{b,d}(t) = \frac{-1 + \sqrt{1 + 2\gamma_{\text{out},d}(t)b}}{\gamma_{\text{out},d}(t)},\\
K_{\text{in}}(t) &= \frac{\exp\left(\frac{1}{2}(b - \hat{\theta}_{b,\text{in}}(t))^2 + \frac{1}{6} \gamma_{\text{in}}(t)\hat{\theta}_{b,\text{in}}(t)^3\right)}{\sqrt{1 + \gamma_{\text{in}}(t)\hat{\theta}_{b,\text{in}}(t)}} \ \text{ with} \ \hat{\theta}_{b,\text{in}}(t) = \frac{-1 + \sqrt{1 + 2\gamma_{\text{in}}(t)b}}{\gamma_{\text{in}}(t)}.
\end{align*}

The only unknown quantities in the above expressions are
$\gamma_{\mathrm{out},w}(t)$, $\gamma_{\mathrm{out},d}(t)$, and
$\gamma_{\mathrm{in}}(t)$, which are obtained from the corresponding
third moments of $R_{\mathrm{out},w}(t)$,
$R_{\mathrm{out},d}(t)$, and $R_{\mathrm{in},1}(t)$.
Their analytic forms are provided in Section B of the supplementary materials.


\subsection{Assessment of $P$-value Approximations under Finite $n$}
\label{sec:cv}
To assess the accuracy of the $p$-value approximations under finite samples, we compare the critical values from the analytical approximations with the critical values obtained via permutations. We set the number of individuals to $n = 200$ and the number of repeated measurements to $\ell = 5$. Under the null hypothesis, data are generated from the multivariate Gaussian distribution with three different dimensions $d = 10$, $50$, and $100$.
Table~\ref{tab:critical_values} presents the critical values at 0.05 significance level for $n_0 = 0.05n$ and $n_0 = 0.1n$ with $n_1=n-n_0$. ``A1'' denotes the analytic critical values without skewness correction, which do not depend on the graph structure. ``A2'' and ``Per'' denote the skewness-corrected analytic critical values and the critical values obtained from 10,000 random permutations, respectively. We construct the similarity graph using 9-MST and use two randomly generated graphs for each configuration to calculate the test statistics and associated critical values.

\begin{table}[h!]
\centering
\caption{Critical values at 0.05 significance level for $n=200$ and $\ell=5$.}
\label{tab:critical_values}

\footnotesize
\setlength{\tabcolsep}{2pt}
\renewcommand{\arraystretch}{0.9}

\begin{minipage}[t]{0.49\textwidth}
\centering
\textbf{(a) $n_0=0.05n$}

\vspace{0.3em}

\begin{tabular}{lccc}
\toprule
   & $Z_{\text{out},w}(t)$
   & $|Z_{\text{out},d}(t)|$
   & $|Z_{\text{in}}(t)|$ \\
\midrule
A1 & 2.986 & 3.032 & 3.032 \\
\bottomrule
\end{tabular}

\vspace{0.5em}

\resizebox{\linewidth}{!}{
\begin{tabular}{lcccccc}
\toprule
\multirow{2}{*}{}
& \multicolumn{2}{c}{$Z_{\text{out},w}(t)$}
& \multicolumn{2}{c}{$|Z_{\text{out},d}(t)|$}
& \multicolumn{2}{c}{$|Z_{\text{in}}(t)|$} \\
\cmidrule(r){2-3}
\cmidrule(r){4-5}
\cmidrule(r){6-7}
& A2 & Per & A2 & Per & A2 & Per \\
\midrule
\multirow{2}{*}{$d=10$}
& 3.192 & 3.281 & 3.033 & 3.029 & 3.074 & 3.103 \\
& 3.209 & 3.343 & 3.033 & 3.024 & 3.073 & 3.082 \\
\midrule
\multirow{2}{*}{$d=50$}
& 3.269 & 3.431 & 3.033 & 3.044 & 3.063 & 3.019 \\
& 3.265 & 3.479 & 3.033 & 3.032 & 3.073 & 3.136 \\
\midrule
\multirow{2}{*}{$d=100$}
& 3.331 & 3.573 & 3.033 & 3.012 & 3.090 & 3.097 \\
& 3.336 & 3.529 & 3.059 & 3.056 & 3.064 & 3.037 \\
\bottomrule
\end{tabular}
}

\end{minipage}
\hfill
\begin{minipage}[t]{0.49\textwidth}
\centering
\textbf{(b) $n_0=0.1n$}

\vspace{0.3em}

\begin{tabular}{lccc}
\toprule
   & $Z_{\text{out},w}(t)$
   & $|Z_{\text{out},d}(t)|$
   & $|Z_{\text{in}}(t)|$ \\
\midrule
A1 & 2.900 & 2.942 & 2.942 \\
\bottomrule
\end{tabular}

\vspace{0.5em}

\resizebox{\linewidth}{!}{
\begin{tabular}{lcccccc}
\toprule
\multirow{2}{*}{}
& \multicolumn{2}{c}{$Z_{\text{out},w}(t)$}
& \multicolumn{2}{c}{$|Z_{\text{out},d}(t)|$}
& \multicolumn{2}{c}{$|Z_{\text{in}}(t)|$} \\
\cmidrule(r){2-3}
\cmidrule(r){4-5}
\cmidrule(r){6-7}
& A2 & Per & A2 & Per & A2 & Per \\
\midrule
\multirow{2}{*}{$d=10$}
& 3.085 & 3.131 & 2.942 & 2.968 & 2.959 & 2.959 \\
& 3.092 & 3.181 & 2.942 & 2.971 & 2.962 & 2.986 \\
\midrule
\multirow{2}{*}{$d=50$}
& 3.135 & 3.277 & 2.942 & 2.938 & 2.956 & 2.936 \\
& 3.142 & 3.270 & 2.942 & 2.933 & 2.960 & 2.953 \\
\midrule
\multirow{2}{*}{$d=100$}
& 3.206 & 3.412 & 2.953 & 2.959 & 2.957 & 2.948 \\
& 3.207 & 3.423 & 2.942 & 2.954 & 2.964 & 2.933 \\
\bottomrule
\end{tabular}
}

\end{minipage}

\end{table}

Overall, our $p$-value approximation performs well, and skewness correction yields more accurate and stable approximations. Across settings, the skewness-corrected critical values closely track the permutation benchmarks, and the agreement further improves when the minimum window size increases (e.g., $n_0 = 0.1n$). Concordance is particularly strong for $|Z_{\text{out},d}(t)|$ and $|Z_{\text{in}}(t)|$, which exhibit negligible differences from permutations. Even for $Z_{\text{out},w}(t)$, whose distribution is highly skewed, the skewness-corrected critical values remain close to the permutation-based critical values. Similar results under the log-normal distribution are provided in Section C in the supplementary materials.


\subsection{Orthogonalized Within-Individual Statistic and Proposed Test} \label{sec:ortho}

While each component, $Z_{\text{out},w}(t)$, $Z_{\text{out},d}(t)$, and $Z_{\text{in}}(t)$, captures a distinct aspect of distributional change, the analytical $p$-value of the combined statistic is complicated by the correlation between $Z_{\text{out},d}(t)$ and $Z_{\text{in}}(t)$. To address this issue, we construct an orthogonalized within-individual edge-count statistic that removes the linear dependence on $Z_{\text{out},d}(t)$, allowing for a simpler joint $p$-value approximation. Based on Lemma~\ref{lemma:evar}, let
\begin{equation}
\varrho_n = \cov(Z_{\text{out},d}(t), Z_{\text{in}}(t)) = 
\frac{\sum_{u=1}^n D_{uu} D_u - \frac{2}{n} |G_{\text{in}}| |G_{\text{out}}|}
{\sqrt{\left( \sum_{u=1}^n D_u^2 - \frac{4 |G_{\text{out}}|^2}{n} \right)
\left( \sum_{u=1}^n D_{uu}^2 - \frac{|G_{\text{in}}|^2}{n} \right)}},
\label{eq:varrho}
\end{equation}
which is invariant with respect to $t$. We first consider the nondegenerate
case $|\varrho_n|<1$. We define the orthogonalized within-individual statistic as
\begin{equation*}
Z_{\text{in}, \perp}(t) = Z_\text{in}(t) - \varrho_n Z_{\text{out},d}(t)
\end{equation*}
with
\begin{align*}
\ep(Z_{\text{in}, \perp}(t)) &= 0, \\
\var(Z_{\text{in}, \perp}(t)) &= \var(Z_\text{in}(t)) - 2\varrho_n \, \cov(Z_{\text{out},d}(t), Z_{\text{in}}(t))+\varrho_n^2 \, \var(Z_{\text{out},d}(t))
\\
&=1-\varrho_n^2.
\end{align*}
\begin{remark}
\label{remark:rho_degenerate}
If $|\varrho_n|=1$, then
$Z_{\text{in}}(t)$ and $Z_{\text{out},d}(t)$ are perfectly linearly
dependent under the permutation null, and hence
$Z_{\text{in},\perp}(t)$ has zero variance. In this degenerate case,
$\tilde Z_{\text{in}}(t)$ is not defined and the within-individual
component provides no additional direction beyond
$Z_{\text{out},d}(t)$. Therefore, the proposed orthogonalized statistic
is considered for $|\varrho_n|<1$; when $|\varrho_n|=1$, one may instead use
the existing between-individual statistic based on
$Z_{\text{out},w}(t)$ and $Z_{\text{out},d}(t)$.
\end{remark}
Then the standardized version of $Z_{\text{in}, \perp}(t)$ is obtained by
\begin{equation} \label{def:tilde}
\tilde{Z}_\text{in}(t) = \frac{Z_{\text{in}, \perp}(t)}{\sqrt{1-\varrho_n^2}}.
\end{equation}

Under Conditions~\ref{con:1}--\ref{con:3}, as $n \to \infty$, the sample correlation $\varrho_n$ defined in (\ref{eq:varrho}) converges to its population counterpart. By construction, $\tilde{Z}_{\text{in}}(t)$ is orthogonal to $Z_{\text{out},d}(t)$, and since Theorem~\ref{thm:gauss} states that $\cov(Z_{\text{in}}(s), Z_{\text{out},w}(t)) = 0$ and $\cov(Z_{\text{out},w}(s), Z_{\text{out},d}(t)) = 0$, we have $\cov(\tilde{Z}_{\text{in}}(s),$ $ Z_{\text{out},w}(t)) = 0$ for all $s,t$. Hence, the processes $\{Z_{\text{out},w}([nu]): 0<u<1\}$, $\{|Z_{\text{out},d}([nu])|: 0<u<1\}$, and $\{|\tilde{Z}_{\text{in}}([nu])|: 0<u<1\}$ converge to mutually independent Gaussian processes as $n \to \infty$. We denote $\{\tilde{Z}_{\text{in}}^*(u): 0 < u < 1\}$ as the limiting Gaussian process corresponding to $\tilde{Z}_{\text{in}}([nu])$. 

To derive the analytical $p$-value approximation, we examine the covariance structure of the proposed statistic. From Theorems~\ref{thm:gauss} and \ref{thm:rho}, $Z_{\text{out},d}(\cdot)$ and $Z_{\text{in}}(\cdot)$ converge to zero-mean Gaussian processes with covariance kernels $\rho_{\text{out},d}(s,t)$ and $\rho_{\text{in}}(s,t)$, respectively, and their cross-covariance satisfies $\cov(Z_{\text{out},d}(s), Z_{\text{in}}(t)) = \varrho_n\,\rho_{\text{out},d}(s,t)$. Then we have
\begin{equation*}
\cov(\tilde{Z}_{\text{in}}(s), \tilde{Z}_{\text{in}}(t))
= \rho_{\text{out},d}(s,t)
= \rho_{\text{in}}(s,t)
\quad \forall s,t.
\end{equation*}
Consequently,
$\cov(\tilde{Z}_{\text{in}}^*(u), \tilde{Z}_{\text{in}}^*(v))
= \rho_{\text{in}}^{*}(u,v)
= \rho_{\text{out},d}^{*}(u,v)$.

This result allows the analytical $p$-value for $|\tilde{Z}_{\text{in}}(t)|$ to be derived in the same way as for $|Z_{\text{out},d}(t)|$ and $|Z_{\text{in}}(t)|$. The analytical $p$-value approximation after skewness correction for $|\tilde{Z}_{\text{in}}(t)|$ is
\begin{equation}
\pr\left( \max_{n_0 \le t \le n_1} |\tilde{Z}_{\text{in}}(t)| > b \right)
\approx 2b \phi(b) \int_{n_0/n}^{n_1/n} \tilde{K}_{\text{in}}(nx) h_{\text{in}}(n,x) \nu\left(b \sqrt{2 h_{\text{in}}(n,x)/n}\right) dx, \label{eq:skew_Ptildein}
\end{equation}
where $h_{\text{in}}(n,x)$ is defined in Remark~\ref{remark:finite} and
\begin{align*}
\tilde{K}_{\text{in}}(t) &= \frac{\exp\left(\frac{1}{2}(b - \hat{\omega}_{b,\text{in}}(t))^2 + \frac{1}{6} \tilde{\gamma}_{\text{in}}(t)\hat{\omega}_{b,\text{in}}(t)^3\right)}{\sqrt{1 + \gamma_{\text{in}}(t)\hat{\omega}_{b,\text{in}}(t)}}, \\
\hat{\omega}_{b,\text{in}}(t) &= \frac{-1 + \sqrt{1 + 2\tilde{\gamma}_{\text{in}}(t)b}}{\tilde{\gamma}_{\text{in}}(t)}, \qquad \tilde{\gamma}_{\text{in}}(t) = \ep(\tilde{Z}^3_\text{in}(t)).
\end{align*}
$\tilde{\gamma}_{\text{in}}(t)$ is computed from the third-order moments of $R_{\text{in},1}(t)$, $R_{\text{out},1}(t)$, and $R_{\text{out},2}(t)$. Specifically,
\begin{equation*}
\ep(\tilde{Z}_\text{in}^3(t)) = \frac{\ep(Z_\text{in}^3(t))-3\varrho_n \, \ep(Z_\text{in}^2(t)Z_{\text{out},d}(t))+3 \varrho_n^2 \, \ep(Z_\text{in}(t)Z_{\text{out},d}^2(t))-\varrho_n^3 \, \ep (Z_{\text{out},d}^3(t))}{\sqrt{(1-\varrho_n^2)^3}}
\end{equation*}
The required third-order moments are given in Section B of the
supplementary materials.

The corresponding critical values for $|\tilde{Z}_{\text{in}}(t)|$ from the same dataset as in Table~\ref{tab:critical_values} are summarized in Table~\ref{tab:critical_values_tilde}. These results are almost the same as the critical values for $|Z_{\text{in}}(t)|$.

\begin{table}[h]
\centering
\caption{Critical values of $|\tilde{Z}_{\text{in}}(t)|$ at 0.05 significance level. $n=200, \ \ell=5$.}
\label{tab:critical_values_tilde}
\footnotesize

\begin{minipage}{0.49\textwidth}
\centering
\subcaption{$n_0 = 0.05n$}
\begin{tabular}{lc}
\toprule
A1 & 3.032 \\
\bottomrule
\end{tabular}

\vspace{0.8em}

\begin{tabular}{lcc}
\toprule
& A2 & Per  \\
\midrule
\multirow{2}{*}{$d=10$} & 3.074 & 3.107 \\
                        & 3.073 & 3.076 \\
\midrule
\multirow{2}{*}{$d=50$} & 3.062 & 3.030 \\
                        & 3.073 & 3.107 \\
\midrule
\multirow{2}{*}{$d=100$} & 3.091 & 3.103 \\
                         & 3.064 & 3.041 \\
\bottomrule
\end{tabular}
\end{minipage}
\hfill
\begin{minipage}{0.49\textwidth}
\centering
\subcaption{$n_0 = 0.1n$}
\begin{tabular}{lc}
\toprule
A1 & 2.942 \\
\bottomrule
\end{tabular}

\vspace{0.8em}

\begin{tabular}{lcc}
\toprule
& A2 & Per \\
\midrule
\multirow{2}{*}{$d=10$}  & 2.959 & 2.950 \\
                         & 2.961 & 2.986 \\
\midrule
\multirow{2}{*}{$d=50$}  & 2.956 & 2.926 \\
                         & 2.960 & 2.950 \\
\midrule
\multirow{2}{*}{$d=100$} & 2.956 & 2.942 \\
                         & 2.964 & 2.941 \\
\bottomrule
\end{tabular}
\end{minipage}

\end{table}

We then propose a final max-type edge-count statistic, which can be expressed as
\begin{equation}
M(t) = \max\left\{ Z_{\text{out},w}(t), |Z_{\text{out},d}(t)|, |\tilde{Z}_{\text{in}}(t)| \right\},
\end{equation}
and the corresponding scan statistic
\begin{equation} \label{finalscanstat}
\max_{n_0 \le t \le n_1}M(t).
\end{equation}

We reject $H_0$ when the scan statistic~(\ref{finalscanstat}) exceeds the critical value at the pre-specified significance level. To obtain this critical value, we compute the corresponding $p$-value $p_{M}$ under the permutation null. The $p$-value is determined from the tail probability of the max-type scan statistic, which can be expressed as
\begin{equation}
\begin{aligned}
&\pr \left(\max_{n_0 \le t \le n_1} M(t) > b \right) \\
& \quad = 1 - \pr \left(\max_{n_0 \le t \le n_1} Z_{\text{out},w}(t) < b \right) \pr \left( \max_{n_0 \le t \le n_1} |Z_{\text{out},d}(t)| < b \right) \pr \left(  \max_{n_0 \le t \le n_1} |\tilde{Z}_{\text{in}}(t)| < b \right).
\end{aligned}
\label{eq:pM}
\end{equation}
The approximations for the constituent scan statistics are directly given (\ref{eq:skew_Pow}), (\ref{eq:skew_Pod}), and (\ref{eq:skew_Ptildein}). The proposed test is summarized in Algorithm $\ref{algo}$.

\begin{algorithm}[h]
    \caption{The new test} \label{algo}
    \begin{algorithmic}[1]
        \REQUIRE A sequence of observations $\{\mathbf{y}_i\}_{i=1,\dots,n}$ and the significance level $\alpha$.
        \ENSURE Reject the null hypothesis $H_{0}$~(\ref{hypo}) if $p_{M}$~(\ref{eq:pM}) is less than $\alpha$. The location of the change-point is estimated using~(\ref{finalscanstat}).
        \STATE Construct the similarity graph $G$ (e.g., 9-MST) on the sequence of observations with repeated measurements.
        \STATE Compute the scan statistic~(\ref{finalscanstat}) based on Lemma~\ref{lemma:evar} and (\ref{def:tilde}).
        \STATE Obtain $p_{M}$ from (\ref{eq:skew_Pow}), (\ref{eq:skew_Pod}), and (\ref{eq:skew_Ptildein}).
    \end{algorithmic}
\end{algorithm}

\subsection{Consistency} \label{subsection:consist}

To establish the consistency of the proposed test and the estimated change-point location as \(n\to\infty\) under a fixed change-point alternative, we first introduce several notations and assumptions. Suppose that there exists a change-point $\tau_n$ such that
\[
\frac{\tau_n}{n}\to\lambda\in(x_0,x_1),
\qquad 
\mathbf{y}_i \sim
\begin{cases}
F_0, & i\le \tau_n,\\
F_1, & i>\tau_n,
\end{cases}
\]
where $F_0\neq F_1$. The individuals $\mathbf{y}_1,\ldots,\mathbf{y}_n$ are independent, while the repeated measurements within each individual may be dependent. We denote the number of between-individual edges crossing the split at $t$ by
\[
R_{\mathrm{out},0}(t)
=
|G_{\mathrm{out}}|
-
R_{\mathrm{out},1}(t)
-
R_{\mathrm{out},2}(t).
\]

\begin{assumption} \label{ass:Rlimit}
We assume that there exist deterministic constants
$r_{\mathrm{out},0}$, $r_{\mathrm{out},1}$,
$r_{\mathrm{out},2}$, $r_{\mathrm{in},1}$, and
$r_{\mathrm{in},2}$ such that
\[
\left(
\frac{R_{\mathrm{out},0}(\tau_n)}{n},
\frac{R_{\mathrm{out},1}(\tau_n)}{n},
\frac{R_{\mathrm{out},2}(\tau_n)}{n},
\frac{R_{\mathrm{in},1}(\tau_n)}{n},
\frac{R_{\mathrm{in},2}(\tau_n)}{n}
\right)^\top
\xrightarrow{p}
\big(
r_{\mathrm{out},0},
r_{\mathrm{out},1},
r_{\mathrm{out},2},
r_{\mathrm{in},1},
r_{\mathrm{in},2}
\big)^\top.
\]
\end{assumption}

\begin{remark}
Assumption~\ref{ass:Rlimit} is a graph-level regularity condition that requires the proportions of the five edge types to stabilize as $n$ increases. For standard geometric similarity graphs including fixed-$k$ $k$-MSTs and $k$-NNGs, weak laws of large numbers for stabilizing graph functionals hold under suitable moment conditions \citep{penrose2003weak}. Individuals remain independent in the repeated-measure setting, making analogous stabilization of the between-individual edge counts natural. Since the number of repeated measurements per individual is fixed, analogous stabilization of the within-individual edge counts is also a natural regularity condition.
\end{remark}

Let $r_{\mathrm{out}} = r_{\mathrm{out},0}+r_{\mathrm{out},1}+r_{\mathrm{out},2}$ and $r_{\mathrm{in}} = r_{\mathrm{in},1} +r_{\mathrm{in},2}$. We define the three limiting change signals as
\begin{align*}
\delta_w
&=
(1-\lambda)r_{\mathrm{out},1}
+\lambda r_{\mathrm{out},2}
-\lambda(1-\lambda)r_{\mathrm{out}}, \\
\delta_d
&=
r_{\mathrm{out},1}
-r_{\mathrm{out},2}
-(2\lambda-1)r_{\mathrm{out}},
\qquad
\delta_{\mathrm{in}}
=
r_{\mathrm{in},1}
-\lambda r_{\mathrm{in}}.
\end{align*}
Based on the limits of Assumption~\ref{ass:limit} in Section~\ref{subsection:null}, let
\[
\varrho
=
\frac{b_6-2b_1b_4}{\sqrt{b_2b_5}},
\qquad
|\varrho|<1,
\]
and
\[
\kappa_w
=
\frac{\delta_w}
{\sqrt{\frac{1}{2}p^2q^2b_3}},
\qquad
\kappa_d
=
\frac{\delta_d}{\sqrt{pqb_2}},
\qquad
\kappa_{\perp}
=
\frac{
\dfrac{\delta_{\mathrm{in}}}{\sqrt{pqb_5}}
-
\varrho\dfrac{\delta_d}{\sqrt{pqb_2}}
}
{\sqrt{1-\varrho^2}}.
\]
Let $\Lambda =
\max \{ \kappa_w,\, |\kappa_d|,\, |\kappa_{\perp}| \}$ and $L_n=\max_{1\le u\le n}(D_u+D_{uu})$. Then we have two additional assumptions.

\begin{assumption} \label{ass:lambda}
$\Lambda>0$.
\end{assumption}

\begin{assumption} \label{ass:L2}
${(L_n^2\log n)}/{n} \xrightarrow{p}0$ as $n \to \infty$.
\end{assumption}

\begin{remark}
Assumption~\ref{ass:lambda} requires a detectable limiting change signal in at least
one graph feature. The three limiting signals can also be written as
\begin{align*}
\delta_w
&=
(1-\lambda)^2r_{\mathrm{out},1}
+\lambda^2r_{\mathrm{out},2}
-\lambda(1-\lambda)r_{\mathrm{out},0},\\
\delta_d
&=
\lambda(1-\lambda)
\left\{
\frac{2r_{\mathrm{out},1}+r_{\mathrm{out},0}}{\lambda}
-
\frac{2r_{\mathrm{out},2}+r_{\mathrm{out},0}}{1-\lambda}
\right\},\\
\delta_{\mathrm{in}}
&=
\lambda(1-\lambda)
\left\{
\frac{r_{\mathrm{in},1}}{\lambda}
-
\frac{r_{\mathrm{in},2}}{1-\lambda}
\right\}.
\end{align*}
These representations show that $\delta_w$ contrasts between-individual
edges connecting observations on the same side of the change-point with
those connecting observations across the change-point.
$\delta_d$ and $\delta_{\mathrm{in}}$ measure the differences in average
between-individual and within-individual degree per individual before and
after the change-point, respectively.
\end{remark}
Assumption~\ref{ass:L2} is naturally satisfied when the graph contains no increasingly dominant hubs with fixed $k$ and a fixed number of repeated measurements.
We denote the estimated change-point location by
\[
\widehat{\tau}_n
=
\min\arg\max_{n_0\le t\le n_1}M(t).
\]

\begin{theorem}
\label{thm:consistency}
Under Assumptions~\ref{ass:limit}--\ref{ass:L2}, the proposed scan statistic is consistent under the fixed change-point alternative in the sense that, for any significance level $0<\alpha<1$,
\[
P\left\{
\max_{n_0\le t\le n_1} M(t) > c_{\alpha,n}
\right\}
\to 1,
\]
where $c_{\alpha,n}$ is the level-$\alpha$ critical value under the
permutation null distribution. The estimated change-point
location satisfies
\[
P\left(
\left|
\frac{\widehat{\tau}_n}{n}-\lambda
\right|>\epsilon
\right)
\to 0
\qquad
\text{for any } \epsilon>0.
\]
\end{theorem}
The proof of Theorem~\ref{thm:consistency} is provided in Section A.4 in the supplementary materials.


\section{Simulation Study} \label{sec:simul}

We examine the performance of the new test under various simulation settings. We consider the observed data generated as follows:
\begin{itemize}
    \item Multivariate Gaussian: $Z_{ij} \overset{\text{ind.}}{\sim} N_d(\theta_{ij}, \omega_i^2 I_d)$.
    \item Multivariate log-normal: $Z_{ij} \overset{\text{ind.}}{\sim} \text{Lognormal}_d(\theta_{ij}, \omega_i^2 I_d)$.
    \item Multivariate Gaussian mixture: $Z_{ij} \overset{\text{ind.}}{\sim} W \cdot N_d(\theta_{ij}, \omega_i^2 I_d) + (1-W) N_d(\theta_{ij}+2, 0.5 \cdot \omega_i^2 I_d), \quad W \sim Bernoulli(0.5)$.
\end{itemize}
with mean vectors $(\theta_{i1}^T, \ldots, \theta_{i\ell}^T) \mid \mu_{ki} \sim N_{d\ell}(\mu_{ki}, \sigma^2 \varrho_k \otimes I_d)$ where $\mu_{ki} = (a_{ki}^T, \ldots, a_{ki}^T)^T$, $a_{ki} \overset{\text{i.i.d.}}{\sim} N_d(\beta_k, \epsilon_k^2 I_d)$ and $\varrho_k = \rho_k \mathbf{1}_\ell \mathbf{1}_\ell^T + (1 - \rho_k) I_\ell,$ and covariance parameter $\omega_i \overset{\text{i.i.d.}}{\sim} U(\nu_{k1}, \nu_{k2})$ for $j = 1, \ldots, \ell$, $i = 1, \ldots, n$. The index $k$ is set to $1$ if the observation time is at or before the true change-point $\tau$, and set to $2$ if it is after $\tau$. Here, $\otimes$ denotes the Kronecker product.

We compare the new method with the generalized edge-count change-point detection
\citep{chu2019asymptotic}, the Fréchet change-point detection
\citep{dubey2020frechet}, and two recent machine learning methods, changeforest and
changekNN \citep{londschien2023random}. To ensure a fair comparison for repeated
measures data, we aggregate the information from repeated measurements using two distance metrics. In the first approach, each individual's repeated measurements are averaged so that the distance is computed as
\begin{equation*}
d(y_{i}, y_{i'}) = d(\tilde{y}_i, \tilde{y}_{i'}) \quad 
\text{where } 
\tilde{y}_i = \frac{1}{\ell} \sum_{j=1}^{\ell} y_{ij}, \ \tilde{y}_{i'} = \frac{1}{\ell} \sum_{j=1}^{\ell} y_{i'j}
\end{equation*}
for any individuals $i \neq i'$, where each $y_{ij}$ is a $d$-dimensional observation. The second approach uses the integrated distance
\begin{equation*} d(y_i, y_{i'}) = \sqrt{\sum_{j=1}^{\ell} \|y_{ij} - y_{i'j}\|^2}, \end{equation*}
treating each repeated measure as a feature and pooling the $\ell$ repeated measurements into $\ell \times d$ dimensions. We denote the generalized edge-count, Fréchet, changeforest, and changekNN methods by $S_b$, $\text{Fréchet}_b$, $\text{changeforest}_b$, and $\text{changekNN}_b$, respectively, where $b=1,2$ indexes the two distances. The new test is denoted $M$.

\begin{table}[h!]
\centering
\caption{Simulation settings for the multivariate Gaussian, log-normal, and Gaussian mixture scenarios with parameters. $\sigma=1$. Each parenthesis shows parameters for $k=1$ (left) and $k=2$ (right). $\mathbf{0}_d$ and $\mathbf{1}_d$ denote the $d$-dimensional zero and one vectors, respectively.}
{\footnotesize
\begin{tabular}{llccccc}
\toprule
\textbf{Distribution} & \textbf{Setting} & $\rho$ & $\boldsymbol{\beta}$ & $\epsilon$ & $\nu_{.1}$ & $\nu_{.2}$ \\
\midrule

\multirow{4}{*}{Gaussian}
& Setting 1 & $(0.2,0.2)$ & $(\mathbf{0}_d,\mathbf{0}_d)$ & $(1,1)$ & $(1,1)$ & $(1.2,1.2)$ \\
& Setting 2 & $(0.1,0.3)$ & $(\mathbf{0}_d,\mathbf{0}_d)$ & $(1,1)$ & $(1,1)$ & $(1.2,1.2)$ \\
& Setting 3 & $(0.2,0.2)$ & $(\mathbf{0}_d,0.3\cdot\mathbf{1}_d)$ & $(1,1)$ & $(1,1)$ & $(1.2,1.2)$ \\
& Setting 4 & $(0.2,0.2)$ & $(\mathbf{0}_d,\mathbf{0}_d)$ & $(1,1.1)$ & $(1,1.1)$ & $(1.1,1.2)$ \\

\midrule

\multirow{4}{*}{Log-normal}
& Setting 1 & $(0.2,0.2)$ & $(\mathbf{0}_d,\mathbf{0}_d)$ & $(1,1)$ & $(1,1)$ & $(1.2,1.2)$ \\
& Setting 2 & $(0.1,0.6)$ & $(\mathbf{0}_d,\mathbf{0}_d)$ & $(1,1)$ & $(1,1)$ & $(1.2,1.2)$ \\
& Setting 3 & $(0.2,0.2)$ & $(\mathbf{0}_d,0.4\cdot\mathbf{1}_d)$ & $(1,1)$ & $(1,1)$ & $(1.2,1.2)$ \\
& Setting 4 & $(0.2,0.2)$ & $(\mathbf{0}_d,\mathbf{0}_d)$ & $(1,1.2)$ & $(1,1.2)$ & $(1.1,1.3)$ \\

\midrule

\multirow{4}{*}{Gaussian mixture}
& Setting 1 & $(0.2,0.2)$ & $(\mathbf{0}_d,\mathbf{0}_d)$ & $(1,1)$ & $(1,1)$ & $(1.2,1.2)$ \\
& Setting 2 & $(0.1,0.4)$ & $(\mathbf{0}_d,\mathbf{0}_d)$ & $(1,1)$ & $(1,1)$ & $(1.2,1.2)$ \\
& Setting 3 & $(0.2,0.2)$ & $(\mathbf{0}_d,0.45\cdot\mathbf{1}_d)$ & $(1,1)$ & $(1,1)$ & $(1.2,1.2)$ \\
& Setting 4 & $(0.2,0.2)$ & $(\mathbf{0}_d,\mathbf{0}_d)$ & $(1,1.1)$ & $(1,1.2)$ & $(1.1,1.3)$ \\

\bottomrule
\end{tabular}}
\label{tab:setting}
\end{table}

\begin{table}[h!]
\centering
\caption{Number of rejections at 5\% significance level (out of 100 replicates) for the multivariate Gaussian data, with the number of detected change-points within [40, 60] in parentheses for $\tau=50$ and [20, 40] in parentheses for $\tau=30$.}
{\footnotesize
\begin{tabular}{llccccccc}
\toprule
& & & \multicolumn{3}{c}{$\tau=50$ (centered)} & \multicolumn{3}{c}{$\tau=30$ (off-center)} \\
\cmidrule(lr){4-6} \cmidrule(lr){7-9}
& & Setting 1 & Setting 2 & Setting 3 & Setting 4 
& Setting 2 & Setting 3 & Setting 4 \\
\midrule
\multirow{9}{*}{$d=40$} 
& $M$ & 2 & 74(56) & 62(51) & 79(73) & 69(60) & 48(42) & 74(65) \\
& $S_1$ & 7 & 16(8) & 74(62) & 49(29) & 19(13) & 49(41) & 52(38) \\
& $S_2$ & 4 & 3(0) & 61(50) & 89(76) & 5(1) & 48(37) & 90(81) \\
& $\text{Fréchet}_1$ & 4 & 0(0) & 12(6) & 9(6) & 2(0) & 20(8) & 9(4) \\
& $\text{Fréchet}_2$ & 1 & 0(0) & 2(0) & 0(0) & 0(0) & 3(0) & 1(0) \\
& $\text{changeforest}_1$ & 3 & 9(1) & 79(69) & 9(1) & 7(0) & 63(46) & 7(1) \\
& $\text{changeforest}_2$ & 3 & 5(2) & 47(35) & 10(2) & 6(1) & 29(19) & 7(2) \\
& $\text{changekNN}_1$ & 9 & 5(0) & 58(38) & 2(0) & 5(0) & 49(32) & 5(0) \\
& $\text{changekNN}_2$ & 4 & 6(0) & 55(35) & 2(0) & 7(1) & 38(23) & 0(0) \\
\midrule
\multirow{9}{*}{$d=50$} 
& $M$ & 6 & 82(72) & 68(58) & 80(62) & 78(68) & 60(52) & 80(71) \\
& $S_1$ & 4 & 23(12) & 77(59) & 47(29) & 30(19) & 61(51) & 52(42) \\
& $S_2$ & 7 & 2(1) & 63(53) & 90(73) & 4(2) & 54(42) & 93(83) \\
& $\text{Fréchet}_1$ & 1 & 2(0) & 16(7) & 4(2) & 2(0) & 25(13) & 7(4) \\
& $\text{Fréchet}_2$ & 0 & 0(0) & 1(0) & 1(1) & 0(0) & 0(0) & 1(0) \\
& $\text{changeforest}_1$ & 5 & 8(3) & 78(67) & 18(8) & 7(1) & 59(49) & 9(2) \\
& $\text{changeforest}_2$ & 8 & 4(0) & 55(39) & 7(1) & 4(1) & 33(27) & 8(2) \\
& $\text{changekNN}_1$ & 12 & 2(0) & 61(35) & 6(0) & 5(1) & 38(26) & 4(0) \\
& $\text{changekNN}_2$ & 3 & 5(0) & 49(26) & 3(0) & 7(0) & 35(22) & 4(0) \\
\bottomrule
\end{tabular}}
\label{tab:mvn_result}
\end{table}

We simulate 100 sequences with $n = 100$ and $\ell = 5$, and set a centered change-point at $\tau=50$ and an off-center change-point at $\tau=30$. The similarity graph $G$ is constructed using the 9-MST. We consider dimensions $d = 40$ and $d = 50$. By varying parameters $\rho, \boldsymbol{\beta}, \epsilon, \nu_{..}, \sigma$, we define four settings. Setting 1 represents the null hypothesis with no change-point. Setting 2 introduces within-individual differences. Setting 3 has the location change between individuals, while Setting 4 changes between-individual scale. The specific parameter values for each setting are listed in Table~\ref{tab:setting}.

\begin{table}[h!]
\centering
\caption{Results for the multivariate log-normal data.}
{\footnotesize
\begin{tabular}{llccccccc}
\toprule
& & & \multicolumn{3}{c}{$\tau=50$ (centered)} & \multicolumn{3}{c}{$\tau=30$ (off-center)} \\
\cmidrule(lr){4-6} \cmidrule(lr){7-9}
& & Setting 1 & Setting 2 & Setting 3 & Setting 4
& Setting 2 & Setting 3 & Setting 4 \\
\midrule
\multirow{9}{*}{$d=40$} 
& $M$ & 2 & 78(67) & 94(76) & 92(78) & 57(47) & 92(84) & 88(76) \\
& $S_1$ & 5 & 4(0) & 88(74) & 91(77) & 3(2) & 82(67) & 92(79) \\
& $S_2$ & 5 & 6(1) & 52(36) & 65(51) & 4(2) & 61(49) & 78(66) \\
& $\text{Fréchet}_1$ & 9 & 5(0) & 27(10) & 37(16) & 2(0) & 13(3) & 17(5) \\
& $\text{Fréchet}_2$ & 7 & 4(0) & 25(7) & 39(14) & 5(0) & 14(2) & 14(3) \\
& $\text{changeforest}_1$ & 5 & 25(13) & 95(94) & 30(14) & 18(4) & 87(72) & 17(4) \\
& $\text{changeforest}_2$ & 2 & 11(2) & 87(79) & 29(12) & 9(2) & 60(46) & 13(2) \\
& $\text{changekNN}_1$ & 4 & 4(0) & 13(0) & 14(0) & 3(0) & 7(0) & 7(0) \\
& $\text{changekNN}_2$ & 5 & 5(0) & 6(0) & 10(0) & 5(1) & 3(0) & 1(0) \\
\midrule
\multirow{9}{*}{$d=50$} 
& $M$ & 4 & 90(81) & 95(85) & 97(85) & 72(48) & 96(91) & 92(87) \\
& $S_1$ & 7 & 6(1) & 82(70) & 89(75) & 6(2) & 82(73) & 87(78) \\
& $S_2$ & 4 & 11(2) & 37(19) & 55(35) & 6(0) & 55(43) & 73(62) \\
& $\text{Fréchet}_1$ & 6 & 3(0) & 14(9) & 35(19) & 1(0) & 8(3) & 12(4) \\
& $\text{Fréchet}_2$ & 4 & 6(0) & 13(6) & 30(12) & 7(4) & 7(2) & 12(5) \\
& $\text{changeforest}_1$ & 11 & 37(14) & 98(94) & 29(15) & 21(2) & 88(79) & 13(3) \\
& $\text{changeforest}_2$ & 8 & 8(3) & 82(75) & 26(10) & 4(0) & 63(46) & 11(1) \\
& $\text{changekNN}_1$ & 6 & 8(0) & 17(0) & 13(0) & 6(0) & 7(0) & 6(0) \\
& $\text{changekNN}_2$ & 4 & 3(0) & 6(0) & 4(0) & 6(0) & 2(0) & 3(0) \\
\bottomrule
\end{tabular}}
\label{tab:ln_result}
\end{table}

\begin{table}[h!]
\centering
\caption{Results for the multivariate Gaussian mixture data.}
{\footnotesize
\begin{tabular}{llccccccc}
\toprule
& & & \multicolumn{3}{c}{$\tau=50$ (centered)} & \multicolumn{3}{c}{$\tau=30$ (off-center)} \\
\cmidrule(lr){4-6} \cmidrule(lr){7-9}
& & Setting 1 & Setting 2 & Setting 3 & Setting 4
& Setting 2 & Setting 3 & Setting 4 \\
\midrule
\multirow{9}{*}{$d=40$} 
& $M$ & 3 & 78(73) & 83(74) & 90(78) & 84(74) & 66(58) & 85(75) \\
& $S_1$ & 3 & 31(22) & 94(90) & 46(37) & 36(19) & 77(67) & 39(33) \\
& $S_2$ & 5 & 6(2) & 81(74) & 99(90) & 8(2) & 60(53) & 98(92) \\
& $\text{Fréchet}_1$ & 7 & 7(1) & 29(10) & 8(1) & 8(2) & 31(8) & 8(3) \\
& $\text{Fréchet}_2$ & 5 & 5(1) & 23(6) & 7(1) & 6(0) & 23(6) & 7(2) \\
& $\text{changeforest}_1$ & 6 & 12(6) & 30(19) & 11(5) & 9(1) & 26(14) & 10(0) \\
& $\text{changeforest}_2$ & 10 & 7(1) & 20(13) & 14(5) & 6(1) & 12(5) & 11(1) \\
& $\text{changekNN}_1$ & 8 & 7(0) & 98(89) & 6(0) & 8(0) & 88(76) & 4(0) \\
& $\text{changekNN}_2$ & 6 & 3(0) & 93(83) & 2(0) & 4(1) & 76(61) & 6(0) \\
\midrule
\multirow{9}{*}{$d=50$} 
& $M$ & 6 & 80(72) & 94(90) & 95(85) & 84(71) & 86(77) & 94(88) \\
& $S_1$ & 6 & 36(25) & 97(92) & 59(46) & 40(29) & 87(81) & 53(43) \\
& $S_2$ & 9 & 8(2) & 89(81) & 100(91) & 7(3) & 72(67) & 99(92) \\
& $\text{Fréchet}_1$ & 3 & 3(0) & 23(13) & 5(1) & 4(1) & 25(13) & 5(1) \\
& $\text{Fréchet}_2$ & 1 & 3(0) & 23(13) & 3(1) & 2(0) & 23(12) & 5(1) \\
& $\text{changeforest}_1$ & 4 & 6(2) & 31(17) & 7(3) & 6(1) & 26(12) & 3(1) \\
& $\text{changeforest}_2$ & 7 & 9(1) & 16(10) & 12(4) & 11(4) & 14(6) & 11(3) \\
& $\text{changekNN}_1$ & 5 & 3(0) & 96(88) & 4(0) & 6(0) & 91(76) & 4(1) \\
& $\text{changekNN}_2$ & 7 & 3(0) & 94(83) & 5(0) & 6(1) & 82(59) & 1(0) \\
\bottomrule
\end{tabular}}
\label{tab:gmm_result}
\end{table}

Table~\ref{tab:mvn_result} reports the number of rejections at the 5\% significance level out of 100 replicates for the multivariate Gaussian data. Under the null (Setting 1), most tests control the type I error rate. For within-individual differences (Setting 2), the proposed test $M$ consistently detects changes with high accuracy, while other methods do not work well. For between-individual differences (Settings 3 and 4), all methods show comparable power; $\text{changeforest}_1$ and $S_1$ attain the highest power in Setting 3, whereas $S_2$ performs best in Setting 4. Since $S_2$ pools the data into $\ell \times d$ dimensions, it requires longer computation time. The Fréchet-based methods perform poorly when the differences before and after the change-point are small. The changeforest and changekNN methods perform well in Setting 3 but poorly in Settings 2 and 4. As is typical for graph-based CPD, power generally improves with dimension. Similar patterns hold for the off-center change-point at $\tau=30$. Overall, $M$ is the only method that reliably detects within-individual differences while remaining effective for between-individual location and scale changes.

Tables~\ref{tab:ln_result} and~\ref{tab:gmm_result} present results for the multivariate log-normal and Gaussian mixture data, respectively. The overall patterns are similar to the Gaussian case. Across both distributions, $M$ performs strongly for within-individual differences (Setting 2) and also performs competitively for between-individual location (Setting 3) and scale (Setting 4) changes.


\section{Real Data Analysis} \label{sec:nyc}
\begin{figure}[h!]
    \centering
    \includegraphics[width=0.65\textwidth]{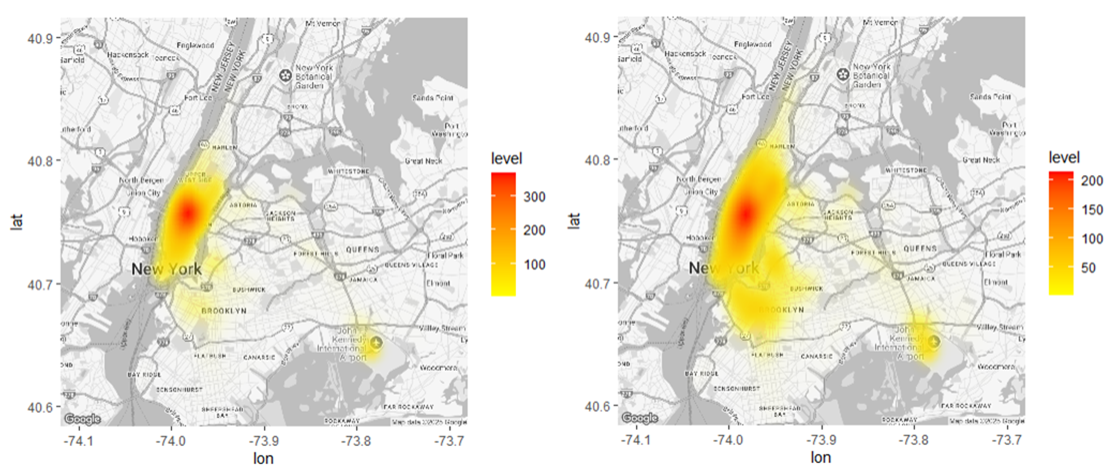}
    \caption{Density heatmaps of taxi drop-offs on Dec 1 (left) and Dec 31 (right), 2015.}
    \label{fig:heatmap}
\end{figure}

To evaluate the proposed method in a real-world setting, we apply it to the NYC Taxi dataset provided by the New York City taxi and Limousine Commission (TLC), publicly available at \url{https://www.nyc.gov/site/tlc/about/tlc-trip-record-data.page}. We use taxi trips picked up at John F. Kennedy (JFK) International Airport and dropped off anywhere in New York City. We first filter all trips that originated from the JFK area. We then divide the New York City area near JFK Airport into a 30 by 30 grid based on longitude and latitude, and construct a matrix of taxi drop-off counts within each cell for each day. Each daily matrix is log-transformed to stabilize large counts. The boundary of this area is set to latitude $40.63$ to $40.66$ and longitude $-73.80$ to $-73.77$. We consider the time period from January 4, 2015 to February 27, 2016, comprising a total of 60 weeks with 420 days. To reflect weekly travel patterns, we treat each day as a repeated measurement and test for a significant change-point across the 60 weekly observations. Figure~\ref{fig:heatmap} shows heatmaps of taxi drop-offs for two example days.

\begin{table}[h!]
\centering
\caption{Detected change-points results in NYC taxi movements. Dates are in 2015 unless otherwise indicated.}
\label{tab:change_points_log}
\footnotesize

\begin{tabularx}{\textwidth}{@{}l l p{0.31\textwidth} X@{}}
\toprule
\textbf{Method} & \textbf{Week} & \textbf{Date} & \textbf{Event} \\
\midrule

$M$
& 5, 12, 43, 50
& Feb 1--7; Mar 22--28; Oct 25--31; Dec 13--19
& Lunar New Year; Spring Break; Halloween Day; Christmas \\
\midrule

$S_1$
& 12, 35, 43
& Mar 22--28; Aug 30--Sep 5; Oct 25--31
& Spring Break; Beginning of Fall Semester; Halloween Day \\
\midrule

$S_2$
& 13, 43
& Mar 29--Apr 4; Oct 25--31
& Spring Break; Halloween Day \\
\midrule

$\text{Fréchet}_1$
& 43
& Oct 25--31
& Halloween Day \\
\midrule

$\text{Fréchet}_2$
& --
& --
& -- \\
\midrule

$\text{changeforest}_1$
& 13, 27, 42
& Mar 29--Apr 4; Jul 5--11; Oct 18--24
& Spring Break; Summer Break; Halloween Day \\
\midrule

$\text{changeforest}_2$
& 19
& May 10--16
& End of Spring Semester \\
\midrule

$\text{changekNN}_1$
& 29
& Jul 19--25
& Summer Break \\
\midrule

$\text{changekNN}_2$
& 6, 12, 45, 53
& Feb 8--14; Mar 22--28; Nov 8--14; Jan 3--9, 2016
& Lunar New Year; Spring Break; Thanksgiving; Christmas \& New Year \\

\bottomrule
\end{tabularx}
\end{table}

\begin{table}[h!]
\centering
\caption{Significant constituent statistics for each detected change-point by $M$.}
\footnotesize
\label{tab:component_pvalues}
\begin{tabular}{c|ccc}
\toprule
Week & $Z_{\text{out},w}(t)$ & $|Z_{\text{out},d}(t)|$ & $|\tilde{Z}_{\text{in}}(t)|$ \\
\midrule
5  & --     & 0.023 & -- \\
12 & 0.001  & 0.009 & -- \\
43 & 0.000  & --    & 0.012 \\
50 & 0.028  & 0.022 & -- \\
\bottomrule
\end{tabular}
\end{table}

The estimated change-points are listed in Table~\ref{tab:change_points_log}. The change-points are estimated using a recursive binary segmentation procedure, which detects all possible change-points at the 5\% significance level. The proposed test $M$ detects four major periods of change, specifically early February (Lunar New Year), late March (Spring Break), late October (Halloween Day), and mid-December (Christmas). These change-points correspond to meaningful events or holidays that plausibly affect NYC taxi movements, but some of these periods are not identified by the competing methods. Similarly, $\text{changekNN}_1$ detects a comparable set of events, while other methods such as $S_1$ and $S_2$ detect changes only within specific periods such as the spring, fall, and Halloween seasons. Table~\ref{tab:component_pvalues} summarizes the $p$-values of the constituent statistics below the 5\% level. Each change-point of $M$ is supported by at least one significant component statistic, showing that the test draws on both between-individual and within-individual variations.


\section{Discussion} \label{sec:disc}

In this paper, we propose a new graph-based change-point framework tailored to a sequence with repeated measurements or local structures. We construct scan statistics that capture both within-individual and between-individual changes and combine them via a max-type strategy. We establish the limiting distributions and consistency of the proposed procedure, and provide accurate, scalable inference through analytical $p$-value approximations with skewness correction, while also offering the permutation-based option when desired.

The proposed test leverages within-individual information when repeated measurements are available. The framework can be extended to accommodate varying numbers of repeated measures across individuals and broader types of within-individual shifts, similar to the location and scale shifts considered in the between-individual setting. Future work may focus on developing more flexible CPD procedures for such settings.

The new procedures identify the most prominent single change-point in a sequence. When multiple changes are present, they can be applied recursively alongside techniques such as binary segmentation, wild binary segmentation, or seeded binary segmentation \citep{fryzlewicz2014wild, kovacs2023seeded}.


\section*{Supplementary Materials}
Supplementary materials provide proofs of theoretical results, analytic derivations for third moments, and additional results on the accuracy of the $p$-value approximation.

\section*{Disclosure Statement}
The authors report there are no competing interests to declare.

\section*{Funding}
This work was supported, in part, by the National Research Foundation of Korea(NRF) grant (RS-2026-25471551,RS-2022-NR068758) and the National IT Industry Promotion Agency(NIPA) grant (RS-2026-25621650), both funded by the Korea government(MSIT).

\setlength{\bibsep}{0pt plus 0.3ex}
\bibliography{reference.bib}

\end{document}